\documentclass[journal,final,letterpaper]{IEEEtran}
\usepackage{amsmath,amssymb,graphicx, subfigure, epsfig}
\usepackage{graphicx}
\usepackage{tikz}
\usetikzlibrary{matrix,positioning}
\usepackage{filecontents}
\usepackage{algorithm2e}
\usetikzlibrary{snakes}
\usetikzlibrary{arrows,decorations,backgrounds}
\newcommand{\LB}{\left(}
\newcommand{\RB}{\right)}
\newcommand{\LSB}{\left[}
\newcommand{\RSB}{\right]}
\newcommand{\LCB}{\left\{}
\newcommand{\RCB}{\right\}}
\newcommand{\lv}{\lvert}
\newcommand{\rv}{\rvert}

\newcommand{\htp}{^{\sf H}}
\newcommand{\tp}{^{\sf T}}

\newcommand{\RR}{{\mathbb R}}
\newcommand{\EE}{{\mathbb E}}

\newcommand{\f}{{\bf f}}
\newcommand{\g}{{\bf g}}
\newcommand{\h}{{\bf h}}

\newcommand{\m}{{\bf m}}

\newcommand{\w}{{\bf w}}
\newcommand{\x}{{\bf x}}

\newcommand{\zerov}{{\bf 0}}

\newcommand{\G}{{\bf G}}

\newcommand{\I}{{\bf I}}

\newcommand{\R}{{\bf R}}

\newcommand{\Cc}{{\cal C}}

\newcommand{\Mc}{{\cal M}}
\newcommand{\Nc}{{\cal N}}

\newtheorem{thm}{Theorem}
\newtheorem{lem}{Lemma}

\newtheorem{prop}{Proposition}

\newtheorem{rem}{Remark}

\newtheorem{app}{Approximation}

\newcommand{\mr}{\mathrm}
\usepackage{color}
\usepackage{acronym}
\usepackage{cite}

\begin{document}

\title{A Stochastic Analysis of Network MIMO Systems}

\author{Kianoush Hosseini,~\IEEEmembership{Member,~IEEE,}
Wei Yu,~\IEEEmembership{Fellow,~IEEE,} \\
and Raviraj S. Adve,~\IEEEmembership{Senior Member,~IEEE}
\thanks{
This paper has been presented in part at the IEEE Global Communications Conference (GLOBECOM) Workshops, Austin, TX, USA, Dec. 2014. This work was supported by the Natural Sciences and Engineering Council (NSERC) of Canada.

K. Hosseini was with the Edward S. Rogers Sr. Department of Electrical and Computer Engineering, University of Toronto, 10 King's College Road, Toronto, Ontario, M5S 3G4, Canada. He is now with Qualcomm Technologies Inc., San Diego, California, 92121, USA (email: kianoush.hosseini@alum.utoronto.ca). 

W. Yu and R. S. Adve are with the Edward S. Rogers Sr. Department of Electrical and Computer Engineering, University of Toronto, 10 King's College Road, Toronto, Ontario, M5S 3G4, Canada (e-mails: weiyu,~rsadve@comm.utoronto.ca). }}

\maketitle

\begin{abstract}
This paper quantifies the benefits and limitations of cooperative communications by providing a statistical analysis of the downlink in network multiple-input multiple-output (MIMO) systems. We consider an idealized model where the multiple-antenna base-stations (BSs) are distributed according to a homogeneous Poisson point process and cooperate by forming disjoint clusters. We assume that perfect channel state information (CSI) is available at the cooperating BSs without any overhead. Multiple single-antenna users are served using zero-forcing beamforming with equal power allocation across the beams. For such a system, we obtain tractable, but accurate, approximations of the signal power and inter-cluster interference power distributions and derive a computationally efficient expression for the achievable per-BS ergodic sum rate using tools from stochastic geometry. This expression allows us to obtain the optimal loading factor, i.e., the ratio between the number of scheduled users and the number of BS antennas, that maximizes the per-BS ergodic sum rate. Further, it allows us to quantify the performance improvement of network MIMO systems as a function of the cooperating cluster size. We show that to perform zero-forcing across the distributed set of BSs within the cluster, the network MIMO system introduces a penalty in received signal power. Along with the inevitable out-of-cluster interference, we show that the per-BS ergodic sum rate of a network MIMO system does not approach that of an isolated cell even at unrealistically large cluster sizes. Nevertheless, network MIMO does provide significant rate improvement as compared to uncoordinated single-cell processing even at relatively modest cluster sizes.

\end{abstract}

\section{Introduction}\label{sec:intro}
\PARstart{N}{}etwork multiple-input multiple-output (MIMO) or coordinated multipoint (CoMP)~\cite{KFV06,GHHSSY10} is often hailed as being capable of completely eliminating intercell interference---a key performance limiting factor in modern cellular networks with densely deployed base-stations (BSs). However, implementing network MIMO systems also comes at considerable costs. In addition to the cost of acquiring channel state information (CSI) as is often mentioned in the literature, this paper points out that cooperating across multiple BSs also introduces a signal penalty, which, together with the inevitable out-of-cluster interference, can have a significant effect on the overall performance of network MIMO systems. The analysis of network MIMO systems that accounts for these effects is a challenging task. The aim of this paper is to quantify the benefits and limitations of cooperative communications by deriving a  \emph{statistical} model of signal and interference strengths in a network MIMO system. 


This paper considers network MIMO systems with \emph{disjoint} clustering in which the BSs within each cluster share data symbols and CSI of all users being served by the cluster. In this architecture, the cooperating BSs in each disjoint cluster form a virtual MIMO system to allow the joint processing of multiple transmissions to multiple users. By cooperating across multiple BSs, network MIMO is able to suppress intercell interference, thereby improving the overall network performance as compared to single-cell processing~\cite{HTHSSV09}. 

This paper analyzes the performance of the linear zero-forcing (ZF) beamforming strategy across multiple BSs. The cooperating BSs are assumed to be connected with delay-free and infinite-capacity backhaul links, and have access to perfect intra-cluster CSI. Making these idealized assumptions allows us to analyze the fundamental limitations of network MIMO systems. We study the cooperation gain of network MIMO systems as a function of two important design factors: $\LB 1\RB$ the loading factor $\eta$, defined as the ratio of the number of scheduled users to the number of BS antennas in each cluster, $\LB 2\RB$ the cluster size, i.e., the number of BSs cooperating in a cluster. 

A system with a large loading factor uses the available spatial dimensions to serve a large number of users, which provides a sum-rate multiplexing gain. However, this comes at the cost of having to use spatial dimensions to suppress intra-cluster interference rather than to provide diversity. In contrast, reducing the loading factor increases the diversity order for each user, at the cost of reduced multiplexing gain. The first objective of this paper is to investigate this tradeoff between the loading factor and the cooperation gain of network MIMO systems.  

The second objective of this paper is to analyze how the per-BS ergodic sum rate of a network MIMO system scales as a function of the cooperating cluster size. We show that surprisingly, despite its ability to eliminate intra-cluster interference, because of both the considerable signal power penalty due to the zero-forcing operation and the inevitable out-of-cluster interference at cluster edge, \emph{a network MIMO system with disjoint clustering can never approach the performance of an isolated cell even as cluster size goes to infinity and perfect CSI acquired without any overhead}. Nevertheless, our analysis does reveal considerable performance improvement of network MIMO as compared to uncoordinated single-cell processing, thus providing useful numerical insights into the design of network MIMO systems in practice.



To carry out a comprehensive analysis of network MIMO performance as a function of these two design parameters, we adopt tools from stochastic geometry~\cite{H12}, accounting for the spatial statistics of BS locations to characterize both the signal strength and inter-cluster interference power, while also accounting for random small-scale fading channel realizations.

\subsection{Related Work}
The benefits of joint processing in network MIMO systems have been extensively studied in the literature. However, most existing analyses of such systems have been carried out either in a single cluster of cooperating BSs~\cite{JTSHSP08,FKV06}, or based on simplified Wyner channel models~\cite{SZS07,SSSP08,JZN12}. Large-system analyses of network MIMO systems have been reported in~\cite{HTC12,RAKL12}. However, it is typically difficult to account for geographic locations of BSs in the analysis. The work in~\cite{HMCD12} studies BS cooperation assuming random user locations in a special case of a two-BS network. Although effective in illustrating the cooperation gains of network MIMO systems, this line of work does not completely account for the spatial statistics of cellular systems. The main objective of this paper is to develop a theoretical framework that takes the random spatial structure of a network and the random channel fading into account.  

Stochastic geometry has been used extensively as a mathematical tool to model the spatial stochastic properties of multi-tier wireless systems with single-cell processing~\cite{DGBA12,DKA13,GDVA14,TDJ15}, cellular systems with interference coordination~\cite{HA13} and cellular systems with non-coherent joint transmission~\cite{TSAJ14}. However, a theoretical analysis of the joint processing scheme of network MIMO under stochastic models is challenging~\cite{TSAJ14,CRP10}, and to the best of the authors' knowledge, not yet available; an area that this paper addresses. Specifically, using tools from stochastic geometry, this paper derives a computationally efficient expression for the per-BS ergodic sum rate of a network MIMO system under ZF beamforming and equal power assignment assumptions. This expression is used to investigate the impact of the loading factor and cluster size on the cooperation gain of network MIMO systems.

Given the available spatial resources at each set of cooperating BSs, the loading factor represents a tradeoff between diversity and spatial multiplexing within each cluster. The loading factor has a significant effect on the overall system performance. For each fixed channel realization, the optimal number of users to schedule can be determined by solving a weighted rate-sum maximization problem~\cite{Z10,HYA13,HTC12}. This paper takes a statistical approach and uses stochastic geometry to determine the optimal loading factor that maximizes the per-BS ergodic sum rate. 

The analysis of this paper
assumes an equal power assignment across the downlink beams. Although sub-optimal in general, equal power allocation is simple to implement, and also allows tractable performance analysis. Assuming equal power assignment, the optimal number of scheduled users that maximizes the spectral efficiency of a non-cooperative massive MIMO system is obtained in~\cite{BLD14}. We note that in addition to providing diversity and intra-cluster interference cancellation, spatial dimensions can also be used for inter-cluster interference nulling if inter-cluster CSI is available. Such possibility has been considered in the literature~\cite{ZCAGH09,MQL15}. The \emph{optimal} use of spatial dimensions for interference nulling is considered in our related work~\cite{HYA15} for the large-scale MIMO system. 

The practical implementation of network MIMO systems relies critically on the feasibility of acquiring and sharing CSI and sharing data across the cooperating BSs. In~\cite{MF09,RCP09}, the impact of imperfect CSI and backhaul constraints on the performance of network MIMO systems is studied. Clearly, increasing the cooperating cluster size increases the CSI acquisition overhead and backhaul communication. The formation of clusters with limited sizes is therefore essential. This paper uses the proposed stochastic geometry analysis to understand how the per-BS ergodic sum rate of network MIMO scales as the cluster size increases. The problem considered in this paper is related to the information theoretical study of uplink network MIMO systems in~\cite{LHA13}. Assuming fixed cluster sizes and perfect CSI, the work in~\cite{LHA13} shows that the spectral efficiency of network MIMO systems is upper bounded as the transmission powers of both serving and interfering clusters grow asymptotically large. Further,~\cite{LHA13} shows that the same conclusion also holds true for a fully cooperative system without CSI at the receivers. This paper considers a different asymptotic regime in the downlink with perfect CSI, where the loading factor and the transmit power per user are kept fixed, while the cluster size grows to infinity.

\subsection{Main Contributions}
The main contributions of this paper are as follows:

\subsubsection{Signal and Inter-Cluster Interference Distributions}
As the first step of the analysis, this paper obtains accurate signal and inter-cluster interference power distributions for a network MIMO system. These distributions make further analysis using stochastic geometry feasible.

\subsubsection{Analysis of Network MIMO Using Stochastic Geometry} Based on the obtained signal and inter-cluster interference power distributions, we derive an accurate expression for the per-BS ergodic sum rate using tools 
from stochastic geometry. This allows us to investigate the cooperation gain of network MIMO systems under different system parameters in a computationally efficient manner.

\subsubsection{Impact of Loading Factor $\eta$}%
We first consider an asymptotic regime where the cluster size grows infinitely large, while $\eta$ is kept fixed. We prove that, when $\eta = 1$, the per-BS ergodic sum rate approaches zero. However, when $\eta < 1$, the per-BS ergodic sum rate is upper bounded by a non-zero constant. This illustrates the crucial impact of the loading factor on the performance of network MIMO systems and motivates the optimization of the per-BS ergodic sum rate as a function of the loading factor. Based on the expression derived, we obtain $\eta^*$, the optimal loading factor that maximizes the per-BS ergodic sum rate. This paper shows that $\eta^* \simeq 0.6$ under a variety of system parameters.

\subsubsection{Impact of Cooperating Cluster Size}
Finally, this paper characterizes the performance of a network MIMO system as a function of the average cluster size. Our analysis reveals that, due to the disparity in the distances between a user and the cooperating BSs, there is a penalty in terms of the received signal power as compared to a conventional cellular system where each user is served only by its closest BS. Further, the performance of network MIMO systems with disjoint clusters is fundamentally constrained by significant out-of-cluster interference even at unrealistically large cluster sizes. Hence, the per-BS ergodic sum rate of a network MIMO system does not approach that of an isolated cell. However, numerical evidence illustrates that network MIMO does provide significant benefit as compared to uncoordinated single-cell processing.

\subsection{Paper Organization}
The remainder of the paper is organized as follows. Section~\ref{sec:sys_signal_model} presents the system model considered in the paper. Section~\ref{sec:dist} presents the statistical characterization of the signal and inter-cluster interference powers. Section~\ref{sec:sg} derives an efficiently computable expression for the achievable per-BS ergodic sum rate. Sections~\ref{sec:loading_factor} and~\ref{sec:cluster_size} employ the per-BS ergodic sum rate expression, respectively, to obtain the optimal loading factor that maximizes the per-BS ergodic sum rate, and to investigate the system performance as a function of the cluster size. Finally, Section~\ref{sec:conc} concludes the paper.

\subsection{Notation}
Matrices, vectors, and scalars are denoted, respectively, by bold capital letters $\G$, bold letters $\g$, and lowercase letters $g$, respectively. The Euclidean norm of a vector is denoted by $\| \cdot \|_2$. The matrix transpose and the Hermitian are indicated by $\LB \cdot \RB \tp$ and $\LB \cdot \RB \htp$, respectively. The $N \times N$ identity matrix is given by $\I_N$. The matrix inverse is denoted by $\G^{-1}$ and $\G^{\dagger} =  \LB \G \htp \G \RB^{-1} \G \htp$ denotes the left pseudo-inverse of $\G$. A multivariate complex Gaussian distribution function with mean $\m$ and covariance matrix $\R$ is indicated by $\Cc\Nc \LB \m,\R\RB$. Further, $\Gamma \LB k,\theta\RB$ represents a Gamma distribution function with scale parameter $k$ and shape parameter $\theta$. Statistical expectation is denoted by $\EE \LB \cdot \RB$, and $X \stackrel{d}{=} Y$ indicates that the two random variables $X$ and $Y$ have the same distribution.

 \section{Stochastic Model of Network MIMO}\label{sec:sys_signal_model}
\begin{figure}[t!]
\centering
\includegraphics[scale = 1.03]{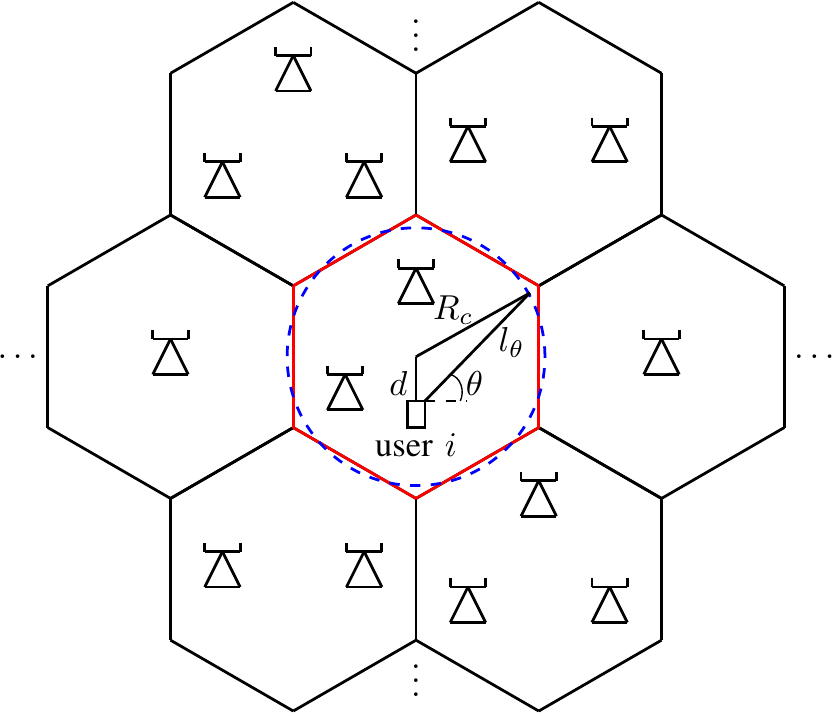}
 \caption{A snapshot of a clustered network MIMO system. Under the stochastic model, the number of BSs in each cluster is a random variable.}
 \label{fig:sys_model}
\end{figure}

This paper characterizes the performance of network MIMO systems in an analytically tractable way. We adopt a stochastic model of the network MIMO that accounts for both random variation of BS locations and the fading in the wireless propagation environment. In the proposed stochastic model, the locations of the BSs are assumed to be a homogeneous Poisson point process (PPP) $\Phi = \LCB \x_i \RCB$ with density $\lambda$, where $\x_i \in \RR^2$ indicates the location of BS $i$. The network MIMO cooperation clusters are formed by grouping BSs into disjoint clusters based on their locations using a fixed hexagonal lattice as shown in Fig.~\ref{fig:sys_model}.\footnote{For analysis purposes, the clustering of BSs and the association of the users to the BS clusters are both location based. In the simulations, discussed in Section~\ref{sec:cluster_size}, channel strength based user association is also considered.} Note that in this stochastic model, the number of BSs in each cluster, and their locations, are all random variables depending on $\Phi$. Let each BS be equipped with $M$ antennas, and for simplicity assume a total power constraint across each cluster with per-BS average power $P_T$. This paper considers an idealized system with perfect knowledge of intra-cluster CSI at the BSs. We ignore the overhead to acquire and exchange CSI and to share users' data symbols across cooperating BSs. 

Assuming $B_l$ cooperating BSs in cluster $l$, and given loading factor $\eta$, $K_l = \eta MB_l$ single-antenna and uniformly distributed users are served within cluster $l$ during each transmission time interval.\footnote{Note that although $\eta$ is identical across clusters, the number of scheduled users might be different; it is dependent on the number of BSs in the cluster.} The network MIMO system is said to be fully-loaded when $\eta = 1$ and partially-loaded when $\eta < 1$.

We further model the wireless fading channel as follows. Let the channel vector between BS $m$ in cluster $j$ and user $i$ in cluster $l$ be denoted by $\g_{ilmj} = \sqrt{\beta_{ilmj}}\h_{ilmj}$, where $\h_{ilmj} \sim \Cc\Nc \LB \zerov,\I_M\RB$, and $\beta_{ilmj} = \LB 1 + \frac{r_{ilmj}}{d_o}\RB^{-\alpha}$. Here, $r_{ilmj}$ indicates the distance between BS $m$ in cluster $j$ and user $i$ in cluster $l$, and $d_o$ is some reference distance. Further, $\alpha > 2$ denotes the path-loss exponent. The collective channel from the $B_l$ cooperating BSs in cluster $l$ to user $i$ is given by $\g_{il} = \LSB \g_{il1l}\tp,\ldots,\g_{ilB_ll} \tp \RSB \tp \in \Cc^{MB_l}$. Likewise, the collective inter-cluster channel from the $B_j$ cooperating BSs in cluster $j$ to user $i$ in cluster $l$ is denoted as $\f_{ilj} = \LSB \g_{il1j} \tp,\ldots,\g_{ilB_j j}\tp \RSB \tp \in \Cc^{MB_j}$.


For simplicity, the total available power in each cluster is equally divided across the downlink beams. Each cluster designs its ZF beams to eliminate intra-cluster interference, while using its excess number of spatial dimensions to provide diversity for its scheduled users. This paper restricts attention to ZF beamforming, because it is relatively easy to implement and to analyze.\footnote{Comparison with minimum mean square error (MMSE) based beamforming scheme is made by simulation in a later section of the paper.} For ZF beamforming, the direction of each beam vector is determined by the projection of the intended channel onto the null space of the subspace spanned by the channels of all other users inside the cluster. Mathematically, the normalized ZF beam assigned to user $i$ in cluster $l$ is given by
\begin{equation}
\w_{il} = \frac{\LB \I_{MB_l} - \G_{-i}\G_{-i}^\dagger \RB \hat{\g}_{il}}{ \Big\| \LB \I_{MB_l} - \G_{-i}\G_{-i}^\dagger \RB \hat{\g}_{il} \Big\|_2},~\forall{i} \in \LCB 1,\cdots,K_l\RCB \nonumber
\end{equation}
where  $\G_{-i} = \LSB \hat{\g}_{1l},\ldots, \hat{\g}_{\LB i-1\RB l},\hat{\g}_{\LB i+ 1\RB l}, \ldots,\hat{\g}_{K_ll}\RSB$, and $\hat{\g}_{il} = \frac{\g_{il}}{\|  \g_{il} \|_2}$.
Note that, due to the orthogonality property of ZF beamforming, the dimension of the beamforming subspace (equivalently, the spatial diversity order) provided to each user in cluster $l$ is 
\begin{equation}
\zeta_l = MB_l \LB 1 - \eta \RB + 1. \label{eq:DoF}
\end{equation}

The received signal at user $i$ is the sum of the intended signal jointly transmitted from a set of cooperating BSs, the inter-cluster interference, and the receiver noise, given by
\begin{align}
y_{il} &= \underbrace{\sqrt{\frac{B_lP_T}{K_l}} \g_{il} \htp \w_{il} s_{il}}_{\text{intended signal}}  + \underbrace{\sum_{j \neq l} \sum_{k = 1}^{K_j} \sqrt{\frac{B_jP_T}{K_j}} \f_{ilj}\htp \w_{kj} s_{kj}}_{\text{inter-cluster interference}} + \underbrace{n_{il}}_{\text{noise}}\nonumber
\end{align}
where $s_{il}$ denotes the complex symbol intended for user $i$ in cluster $l$ such that $\EE \LSB \lv s_{il} \rv^2\RSB = 1$, and $n_{il} $ denotes the circularly symmetric complex additive white Gaussian noise with variance $\sigma^2$. 
The signal-to-interference-and-noise ratio (SINR) of user $i$ located at distance $d$ from the center of cluster $l$ is then expressed as
\begin{equation}
\mr{\gamma}_{il,d} = \frac{\rho \lv \g_{il}\htp \w_{il}\rv^2}{\sum_{ j \neq l} \sum_{k} \rho \lv \f_{ilj}\htp \w_{kj}\rv^2  + 1} \label{eq:sinr}
\end{equation}
where $\rho = \frac{P_T}{\eta M\sigma^2}$ denotes the signal-to-noise ratio (SNR). The ergodic rate of user $i$ is given by

\begin{equation}
R_{il,d} = \EE_{\Phi,\h} \LSB \log_2 \LB 1 + \frac{\gamma_{il,d}}{\Gamma} \RB \RSB  \label{eq:cap}
\end{equation}
where $\Gamma$ is the SNR gap determined by the modulation and coding scheme used in a system. We denote the set of the serving BSs within cluster $l$ by $\Phi_S$, and the set of interfering BSs by $\Phi_{I} = \Phi \setminus \Phi_S$. 

For analytical tractability, the typical hexagonal cluster is approximated by a circular cluster of radius $R_c$ with the same area as shown in Fig.~\ref{fig:sys_model} when computing both signal and interference strengths. 

\section{Signal and Interference Power Distributions}
\label{sec:dist}
As can be seen from~\eqref{eq:sinr}, the signal power of each user is proportional to the power of the intended channel projected onto the beamforming subspace. Similarly, the interference power created by each beam in an interfering cluster is proportional to the power of the inner-product of the inter-cluster channel and the beam vector. For isotropic channels, i.e., channels comprising only independent and identically distributed (i.i.d.) entries, the distributions of the signal and interference powers have been derived previously~\cite{FM90,CKA09}. In network MIMO systems, each composite channel is subject to multiple path-loss parameters from multiple BSs and is, therefore, non-isotropic. The analysis of these non-isotropic distributions is far more challenging. This paper utilizes a moment-matching technique proposed in~\cite{SHCS14,HWKS11} to approximate the signal and interference distributions in a network MIMO system as isotropic distributions. The technique of~\cite{SHCS14,HWKS11} applies only to systems with \emph{deterministic BS locations}. This paper proposes further approximations that allow us to derive signal and interference distributions that are amenable to further analysis using stochastic geometry to account for random BS locations.

\subsection{Preliminary: Projection of Isotropic Channels}
We first present some preliminary facts for the case when the channels are isotropic. Let $\x = \LSB x_1,\ldots,x_{N} \RSB \in \Cc^{N}$ be a complex $N$-dimensional isotropic vector. Assuming that each entry is distributed as $\Cc\Nc \LB 0,\beta\RB$, then $\x \htp \x$ is the summation of $N$ i.i.d. exponentially distributed random variables; we have 
\begin{equation}
\x \htp \x \sim \Gamma \LB N,\beta \RB. \nonumber
\end{equation}

It is important to note that, when the vector $\x$ is isotropic, each of the $N$ spatial dimensions adds one to the shape parameter of the power distribution. Further, since $\x$ is isotropic, when projected onto an $s$-dimensional subspace, its power is distributed as~\cite{M82}
\begin{equation}
\lv \x\htp \w\rv^2 \sim \Gamma \LB s,\beta\RB. \nonumber 
\end{equation}
Again, each of the $s$ spatial dimensions contributes one to the shape parameter of the power distribution. The scale parameter remains the same.

These observations are used in the following subsection to approximate the distributions for non-isotropic channels.

\subsection{Approximate Distribution of Non-Isotropic Channels} 
In a network MIMO system, the intended channel and the interference channel strengths are given, respectively, by 
\begin{align}
\g_{il}\htp\g_{il} &= \sum_{b = 1}^{B_l} \beta_{ilbl}\h_{ilbl}\htp\h_{ilbl} \nonumber \\
 \f_{ilj}\htp\f_{ilj} &= \sum_{m = 1}^{B_j} \beta_{ilmj}\h_{ilmj}\htp\h_{ilmj} \nonumber
\end{align}
as summations of $B_l$ and $B_j$ independent, but non-identically distributed, Gamma random variables; the exact distributions of the channel strengths therefore involve complex expressions. To make the distributions more amenable to analysis, we use the \emph{second-order moment matching} technique~\cite{SHCS14,HWKS11}.

\begin{lem}[Gamma Second-Order Moment Matching] Let $\LCB X_i\RCB_{i = 1}^{m}$ be a set of $m$ independent random variables such that $X_i \sim \Gamma \LB k_i,\theta_i \RB$. Then,
$Y = \sum_{i} X_i$ has the same first and second order statistics as the Gamma random variable with the shape and scale parameters given by
\begin{equation}
k = \frac{\LB \sum_i k_i \theta_i \RB^2}{\sum_i k_i \theta_i^2}~~\text{and}~~\theta = \frac{\sum_i k_i \theta_i^2}{\sum_i k_i \theta_i }. \label{eq:shape_scale}
\end{equation}
\label{mom_match}
\end{lem}

\begin{app}\label{mom_match_app}
Based on Lemma~\ref{mom_match}, we approximate the distributions of the channel strengths as $\g_{il}\htp \g_{il} \sim \Gamma \LB k_{il},\theta_{il}\RB$ and $\f_{ilj} \htp \f_{ilj} \sim \Gamma \LB k_{ilj},\theta_{ilj}\RB$ wherein 
\begin{align}
k_{il} &= M \frac{\LB \sum_{b = 1}^{B_l} \beta_{ilbl} \RB^2}{ \sum_{b = 1}^{B_l} \beta_{ilbl}^2},~\theta_{il} = \frac{\sum_{b = 1}^{B_l} \beta_{ilbl}^2} {\sum_{b = 1}^{B_l} \beta_{ilbl}} \label{eq:intended_chan} \\
k_{ilj} &= M \frac{\LB \sum_{m = 1}^{B_j} \beta_{ilmj} \RB^2}{ \sum_{m = 1}^{B_j} \beta_{ilmj}^2},~\theta_{ilj} = \frac{\sum_{m = 1}^{B_j} \beta_{ilmj}^2} {\sum_{m = 1}^{B_j} \beta_{ilmj}}. \label{eq:intf_chan}
\end{align}
\end{app}

Having obtained approximate distributions of the channel magnitudes, to facilitate the characterization of the power distributions when channels are projected onto the beamforming subspace, we now further assume that $\g_{il}$ and $\f_{ilj}$ are isotropic based on the scale parameters of $\g_{il} \htp \g_{il}$ and $\f_{ilj}\htp \f_{ilj}$, respectively, as follows: 

\begin{app}
The intended channel distribution is approximated as
\begin{equation}
\g_{il} \sim \Cc\Nc \LB \zerov,\theta_{il}\I_{MB_l}\RB. \nonumber
\end{equation}
Likewise, the interference channel distribution between cluster $j$ and user $i$ in cluster $l$ is approximated as 
\begin{equation}
\f_{ilj} \sim \Cc\Nc \LB \zerov,\theta_{ilj} \I_{MB_j}\RB. \nonumber
\end{equation}
\label{isotropic}
\end{app}
However, unlike the truly isotropic case, we assume that each spatial dimension only contributes a fraction to the shape parameter of the associated power distributions~\cite{SHCS14}. In particular, note from~\eqref{eq:intended_chan} and~\eqref{eq:intf_chan} that, in contrast to the true isotropic case, here we have
\begin{equation}
 k_{il} \leq MB_l~\text{and}~ k_{ilj} \leq MB_j \nonumber
 \end{equation}
 with equality if the two vectors were isotropic. Therefore, it can be assumed that each spatial dimension adds, respectively, $\frac{k_{il}}{MB_l}$ and $\frac{k_{ilj}}{MB_j}$ to the shape parameter of the signal and interference power distributions.

Based on this assumption, the approximated signal and interference power statistics under ZF beamforming can now be obtained. The signal power is the power of $\g_{il}$ when projected onto the beamfoming subspace. If the dimension of the beamforming subspace is $\zeta$, then the shape parameter of the signal power distribution is given by $\frac{k_{il}\zeta}{MB_l}$. 

The interference power created at user $i$ by the transmission of a single beam $\w_{nj}$ is the power of the inner-product between $\f_{ilj}$ and $\w_{nj}$. Note that $\f_{ilj}$ and $\w_{nj}$ are independent. Thus, the shape parameter of the interference power distribution is $\frac{k_{ilj}}{MB_j}$. The following lemma presents the approximate signal and interference power distributions~\cite{SHCS14}:



\begin{lem}
\label{sig_intf_dist}
In a network MIMO system employing ZF beamforming in each cluster, under Approximations~\ref{mom_match_app} and~\ref{isotropic} and the aforementioned assumptions, the distribution of the signal power at user $i$ in cluster $l$ and the distribution of the interference power created by transmission of beam $n$ in cluster $j$ at user $i$ in cluster $l$ are given, respectively, as\begin{align}
&\lv \g_{il}\htp \w_{il} \rv^2 \sim \Gamma \LB \frac{k_{il}\LB MB_l \LB 1 - \eta \RB + 1\RB}{MB_l}, \theta_{il}\RB \label{eq:signal_dist} \\
&\lv \f_{ilj}\htp \w_{nj}\rv^2 \sim \Gamma \LB \frac{k_{ilj}}{MB_j},\theta_{ilj}\RB \label{eq:intf_dist}
\end{align}
with $k_{il}$, $\theta_{il}$, $k_{ilj}$, and $\theta_{ilj}$ defined in~\eqref{eq:intended_chan} and~\eqref{eq:intf_chan}.
\end{lem}

Although accurate for any fixed set of BS locations, the distribution functions presented in Lemma~\ref{sig_intf_dist} still do not lead to a tractable analysis of network MIMO with random BS locations. 
To facilitate an analysis that takes the spatial statistics of the BS deployment into account, we must further approximate the signal and interference power distributions.

\subsection{Approximate Signal Power Distribution}

Based on Lemma~\ref{mom_match}, the signal power $\lv \g_{il} \htp \w_{il} \rv^2$ presented in~\eqref{eq:signal_dist} has the same first and second order statistics as $\sum_{b = 1}^{B_l} \beta_{ilbl} \kappa_b $ where
\begin{equation}
\kappa_b \sim \Gamma \LB \frac{MB_l \LB 1 - \eta \RB + 1}{B_l},1 \RB. \label{eq:kappa}
\end{equation}
We can therefore further approximate the signal power of user $i$ in cluster $l$ as follows:

\begin{app} \label{newapp}
The signal power of user $i$ in cluster $l$ can be approximated by
\begin{equation}
\lv \g_{il}\htp \w_{il}\rv^2 \stackrel{d}{=} \sum_{b = 1}^{B_l} \beta_{ilbl} \kappa_b  \label{eq:sig_app1}
\end{equation}
\label{sig_mom_match1}
where $\kappa_b$ is given by~\eqref{eq:kappa}.
\end{app}

Using this approximation, the signal power is decomposed into summation of $B_l$ terms, where the distribution of each term depends on the distance between user $i$ and only one of the serving BSs in the cluster. This greatly facilitates further stochastic geometry analysis. However, the distribution of $\kappa_b$ in this expression is dependent on the number of serving BSs $B_l$, which is a random variable. To make the analysis tractable, we replace $B_l$ by the average number of BSs in the cluster $\bar{B}$, and further approximate the signal power as follows:

\begin{app}
The signal power of user $i$ in cluster $l$ is given by
\begin{equation}
\lv \g_{il}\htp \w_{il}\rv^2 \stackrel{d}{=} \sum_{b = 1}^{B_l} \beta_{ilbl} \tilde{\kappa} \nonumber
\end{equation}
 wherein
\begin{equation}
\tilde{\kappa} \sim \Gamma \LB \frac{M\bar{B}  \LB 1 - \eta \RB + 1}{\bar{B}},1 \RB. \label{eq:sig_app2}
\end{equation}
\label{sig_mom_match2}
\end{app}

The next subsection derives a tractable distribution function for the inter-cluster interference power.

\begin{figure*}[!t]
\begin{subequations}
\label{eq:sub_eqn}
\begin{eqnarray}
&&\hspace{-0.5cm}R_{{cell}} =  \eta M \int_{0}^{R_c} f_{d}\LB d\RB \times \nonumber \\
& &\LSB \int_{0}^{\infty} \frac{e^{-z\Gamma}}{z} \exp \LB -\lambda \int_{0}^{2\pi} \LB \Psi_{\mr{I}} \LB \theta \RB - \Psi_{\mr{II}} \LB \theta \RB \RB  \mr{d} \theta\RB   \LB 1 -  \exp \LB -\lambda \int_{0}^{2\pi} \LB \Upsilon_{\mr{I}} \LB \theta \RB - \Upsilon_{\mr{II}} \LB \theta \RB \RB  \mr{d} \theta\RB \RB \mr{d}z \RSB \mr{d}d  \\
&&\hspace{-0.5cm}\Psi_{\mr{I}} \LB \theta \RB  =  \frac{d_o^2\LB 1 +\frac{ l_{\theta}}{d_o} \RB^2}{2} \LSB {}_2 F_1 \LB \eta M,\frac{-2}{\alpha};1 - \frac{2}{\alpha};-\rho z \Gamma \LB 1 + \frac{l_{\theta}}{d_o} \RB^{-\alpha} \RB - 1 \RSB \\
&&\hspace{-0.5cm}\Psi_{\mr{II}} \LB \theta \RB  =  d_o^2\LB 1 + \frac{l_{\theta}}{d_o} \RB \LSB {}_2 F_1 \LB \eta M,\frac{-1}{\alpha};1 - \frac{1}{\alpha};-\rho z \Gamma \LB 1 + \frac{l_{\theta}}{d_o} \RB^{-\alpha} \RB - 1 \RSB \\
&&\hspace{-0.5cm}\Upsilon_{\mr{I}} \LB \theta \RB  =  \frac{d_o^2\LB 1 + \frac{l_{\theta}}{d_o} \RB^2}{2} \LSB 1 - {}_2F_1 \LB \varpi,\frac{-2}{\alpha};1 - \frac{2}{\alpha};-\rho z \LB 1 + \frac{l_{\theta}}{d_o} \RB^{-\alpha} \RB \RSB - \frac{d_o^2}{2} \LSB 1 - {}_2F_1 \LB \varpi,\frac{-2}{\alpha};1 - \frac{2}{\alpha};-\rho z \RB \RSB  \\
&&\hspace{-0.5cm}\Upsilon_{\mr{II}} \LB \theta \RB  =  d_o^2\LB 1 + \frac{l_{\theta}}{d_o} \RB \LSB 1 - {}_2F_1 \LB \varpi,\frac{-1}{\alpha};1 - \frac{1}{\alpha};-\rho z \LB 1 + \frac{l_{\theta}}{d_o} \RB^{-\alpha} \RB \RSB - d_o^2\LSB 1 - {}_2F_1 \LB \varpi,\frac{-1}{\alpha};1 - \frac{1}{\alpha};-\rho z \RB \RSB \\
\hline \nonumber
\end{eqnarray}
\end{subequations}
\end{figure*}

\subsection{Approximate Interference Power Distribution}
Since there is no coordination across clusters, interference terms from different clusters are mutually independent. Hence, we focus on the interference power produced by cluster $j$ at user $i$ in cluster $l$, denoted by 
\begin{equation}
I_{il}^j = \sum_{k = 1}^{K_j} \lv \f_{ilj} \htp \w_{kj} \rv^2. \nonumber 
\end{equation}
Since the ZF beam vectors are not necessarily mutually orthogonal, $I_{il}^j$ is a summation of $K_j$ identically distributed, but dependent, Gamma random variables. As a result, obtaining an analytic expression of the distribution of $I_{il}^j$ is difficult. To tackle this problem, using an approach similar to that taken in~\cite{SHCS14,CKA09,HWKS11,DKA13}, we assume that the ZF beams designed in each interfering cluster are mutually orthogonal. Under this assumption, $I_{il}^j$ is a summation of $K_j$ i.i.d. Gamma random variables where each term is distributed as in~\eqref{eq:intf_dist}; therefore
\begin{equation}
I_{il}^j \sim \Gamma \LB \eta k_{ilj},\theta_{ilj} \RB. \label{eq:intf_app1}
\end{equation}
 
Similar to the signal power distribution, both the shape and scale parameters of the distribution function in~\eqref{eq:intf_app1} are jointly dependent on the distances between user $i$ and the set of BSs in cluster $j$. Based on Lemma~\ref{mom_match}, the inter-cluster interference power produced by cluster $j$ has the same first and second order statistics as $\sum_{m = 1}^{B_j} \beta_{ilmj} \psi_{ilmj}$ where
\begin{equation}
\psi_{ilmj} \sim \Gamma \LB \eta M,1\RB. \label{eq:psi}
\end{equation}
Thus, $I_{il}^j$ can further be approximated as follows:

\begin{app}
The aggregate interference power created by cluster $j$ at user $i$ in cluster $l$ can be written as 
\begin{equation}
I_{il}^{j} \stackrel{d}{=} \sum_{m = 1}^{B_j} \beta_{ilmj} \psi_{ilmj} \label{eq:intf_app2}
\end{equation}
where $\psi_{ilmj}$ is given in~\eqref{eq:psi}.
\label{intf_app2}
\end{app}

According to~\eqref{eq:intf_app2}, the interference power imposed by cluster $j$ is a summation of $B_j$ terms where each term depends only on the distance between a single BS in cluster $j$ and user $i$. Hence, the cluster index $j$ can be dropped. In fact, the aggregate interference power $\sum_{j\neq l} I_{il}^j$ in a network MIMO system is equivalent, in distribution, to that of a system where each BS $m \in \Phi_I $ independently produces interference whose power is distributed as $\beta_{ilm} \psi_{ilm}$, i.e.
\begin{equation}
\sum_{j \neq l}\sum_{m = 1}^{B_j} \beta_{ilmj} \psi_{ilmj} \stackrel{d}{=} \sum_{m \in \Phi_I} \beta_{ilm} \psi_{ilm}.
\end{equation}

The obtained signal and interference power distribution functions in Approximations~\ref{sig_mom_match2} and~\ref{intf_app2} can now be readily used to analyze the per-BS ergodic sum rate of a network MIMO system.

\section{Stochastic Geometry Analysis of Network MIMO Systems }\label{sec:sg}

In this section, we assume that the BSs are randomly distributed and use the approximate signal and interference distributions to characterize the per-BS ergodic sum rate of a network MIMO system, while accounting for the spatial statistics of BS deployment and random small-scale fading channel realizations. The following theorem provides the per-BS ergodic sum rate expression as a function of key system parameters such as cooperating cluster size $\bar{B},~$SNR $\rho$, loading factor $\eta$, the number of antennas per BS $M$, BS deployment density $\lambda$, SNR gap $\Gamma$, and channel model parameters such as path-loss exponent $\alpha$ and reference distance $d_o$, using theoretical tools from stochastic geometry.

\begin{thm}

Under Approximations~\ref{mom_match_app}-\ref{intf_app2}, the per-BS ergodic sum rate of a network MIMO system with ZF beamforming and equal power assignment in each cluster is given by~\eqref{eq:sub_eqn} in nats per second per Hertz. In~\eqref{eq:sub_eqn} , $\varpi$ indicates the shape parameter of the Gamma distribution in~\eqref{eq:sig_app2} given by
\begin{equation}
\varpi = \frac{M\bar{B} \LB 1 - \eta \RB + 1}{\bar{B}} \nonumber
\end{equation}
and ${}_2 F_1 \LB a,b;c;z\RB$ denotes the Gauss hypergeometric function given by
\begin{equation}
{}_2 F_1 \LB a,b;c;z\RB = \frac{\Gamma \LB a \RB}{\Gamma\LB b\RB \Gamma \LB c - b\RB} \int_0^1 \frac{s^{b - 1} \LB 1 - s \RB^{c - b - 1}}{\LB 1 - sz \RB^a}\mr{d}s. \nonumber
\end{equation}
The Gamma function is given as $\Gamma \LB x \RB = \int_{0}^{\infty} t^{x - 1}e^{-t}\mr{d}t$. The probability density function of $d$ is\begin{equation}
f_d \LB d\RB = \left\{\begin{array}{l l} \frac{2d}{R_c^2}, & \quad \text{if $d\in \LSB 0,R_c\RSB$} \\ 0,  & \quad \text{otherwise} \end{array}\right.. \label{eq:dist_dist}
\end{equation}

\label{ergodic_cap}
\end{thm}

 \begin{IEEEproof}
See Appendix~\ref{sec:app_cap}.
\end{IEEEproof}

\begin{rem}
By expressing the hypergeometric functions in~\eqref{eq:sub_eqn} using upper incomplete beta functions as suggested in Remark $1$ of~\cite{ZH14}, the per-BS ergodic sum rate expression can be evaluated efficiently.
\end{rem}

\begin{rem}
By dropping the outer integral, the expression in~\eqref{eq:sub_eqn} yields the ergodic rate of a user as a function of its distance to the cluster center. The user ergodic rate is an important performance metric in cellular networks and can be exploited to investigate the quality-of-service of the users at various locations inside a cooperating cluster.
\end{rem}

The expression in~\eqref{eq:sub_eqn} completely and efficiently characterizes the per-BS ergodic sum rate of a network MIMO system as a function of key system parameters such as transmit power, loading factor, BS deployment density, and cooperating cluster size. Although the expression involves a triple integral, it is nevertheless much faster to evaluate than performing system-level simulations. This analytic result allows us to obtain the optimal loading factor that maximizes the per-BS ergodic sum rate of a network MIMO system. In addition, we can investigate how much gain, in terms of the per-BS ergodic sum rate, can be achieved by increasing the average cluster size in network MIMO systems. These results are presented in the following two sections.

\section{Network MIMO Performance as a Function of Loading Factor}\label{sec:loading_factor}
Using the proposed analytical framework, this section investigates the effect of the loading factor on the per-BS ergodic sum rate of network MIMO systems. To confirm the analytical derivation (i.e.,~\eqref{eq:sub_eqn}), we also provide numerical results using the following system parameters. 



The density of the PPP corresponding to the locations of the BSs is set to $\lambda = \frac{1}{\pi 500^2}$ $\mr{m}^{-2}$ , i.e., on average, the inter-BS distance is $1$km. In the numerical simulations, we consider $19$ hexagonal cooperating clusters, each of area $\frac{\bar{B}}{\lambda}$, and focus on the performance of the center cluster. BSs are assumed to be equipped with $M = 5$ antennas, and their maximum available power is set to $43$dBm. The downlink transmissions take place over the shared spectrum of bandwidth $W = 20$MHz with universal frequency reuse factor of one; transmissions from different clusters interfere with each other. 

During each transmission time interval, $K_j = \eta M B_j$ single-antenna users are randomly selected among users located within the cluster $j$. The small-scale channel fading coefficients are generated according to a Rayleigh distribution and assumed to be independent across both the transmit antennas and time. The numerical results are averaged over both spatial topologies and small-scale fading realizations. The system parameters are summarized in Table~\ref{table:parameters}. 

\begin{table}[!t]
\caption{System Design Parameters}
\centering
    \begin{tabular}{|c| c|}
    \hline
    BS density & $\lambda = 1/\pi 500^2$ m$^{-2}$\\ \hline
    Total bandwidth & $W = 20$ MHz \\ \hline
    BS max. available power & $43$ dBm \\ \hline
    Background noise & $N_o = -174$ dBm/Hz \\ \hline
    Noise figure & $N_f = 9$ dB \\ \hline
    SNR gap & $\Gamma = 3$ dB \\ \hline
    Path-loss exponent & $\alpha =3.76$ \\ \hline
    Reference distance & $d_o = 0.3920$ m \\ \hline
    BS antennas & $M = 5$ \\ \hline
    
         \end{tabular}
    \label{table:parameters}
\end{table}

\subsection{Per-BS Ergodic Sum Rate with $\eta = 1$ and $\eta < 1$}
\begin{figure}[t!]
\centering
\includegraphics[scale = 1.03]{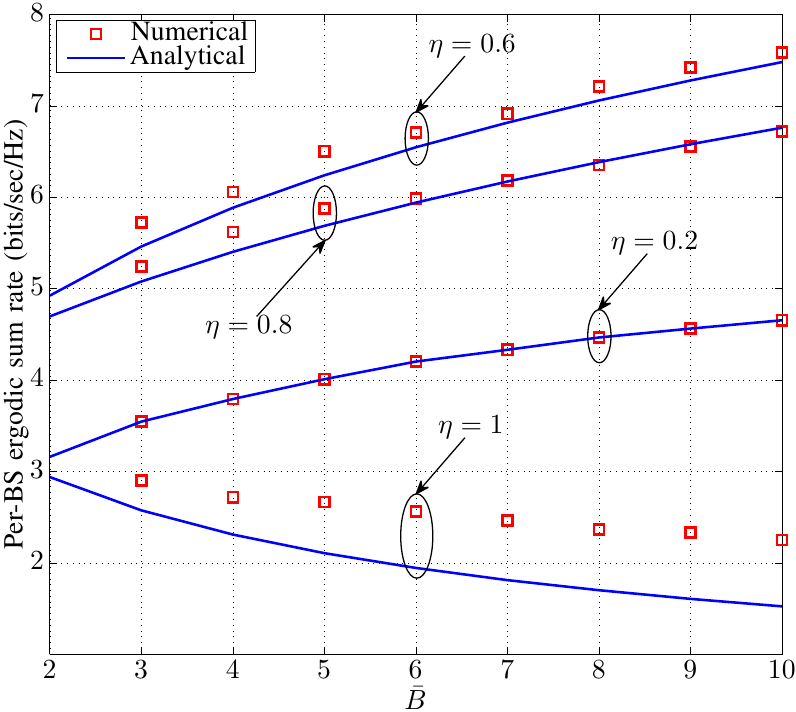}
\caption{Per-BS ergodic sum rate as a function of $\bar{B}$ for various values of $\eta$.}
\label{fig:per_cell}
 \end{figure}

First, we evaluate the per-BS ergodic sum rate of a network MIMO system with various average cluster sizes $\bar{B}$ and loading factors $\eta$ in Fig.~\ref{fig:per_cell}. As can be seen from this figure, the analytical and numerical results match under different values of $\eta$ and $\bar{B}$. This confirms the accuracy of the approximations made in Section~\ref{sec:dist}.

Further, for any given $\bar{B}$, the system performance improves as $\eta$ increases from $0.2$ to $0.6$; however, the system performance degrades as $\eta$ is further increased. This suggests that there exists an optimal loading factor that maximizes the per-BS ergodic sum rate. Another important observation that can be drawn from Fig.~\ref{fig:per_cell} is that for various values of $\eta < 1$, the per-BS ergodic sum rate increases as the average cluster size grows. However, in a fully-loaded system with $\eta = 1$, the per-BS ergodic sum rate actually decreases as the average cluster size increases. In the following, we first numerically obtain the optimal loading factor, then discuss the surprising phenomenon in the $\eta = 1$ case in some detail.

\subsection{Optimal Loading Factor}
\begin{figure} 
\centering
\includegraphics[scale = 1.03]{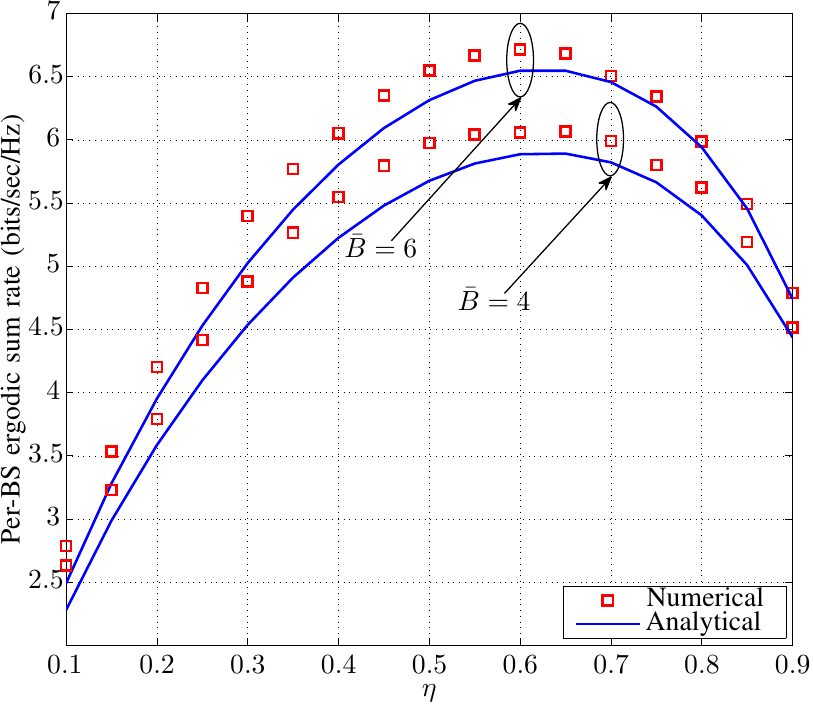}
\caption{Per-BS ergodic sum rate as a function of $\eta$ for various values of $\bar{B}$.}
\label{fig:sum_rate5}
 \end{figure}

We consider the optimization of loading factor to maximize the per-BS ergodic sum rate. Fig.~\ref{fig:sum_rate5} plots the per-BS ergodic sum rate as a function of $\eta$ for $\bar{B} = 4$ and $\bar{B} = 6$. As both the numerical and analytical results illustrate, the per-BS ergodic sum rate reaches a peak when $\eta \simeq 0.6$. We further observe that $\eta^*$ remains the same for different choices of $M$, $\bar{B}$, and $\lambda$ as well. 

It is important to note that since CSI acquisition overhead is not accounted for, the obtained $\eta^*$ upper bounds the optimal loading factor in a practical implementation of a network MIMO system. In particular, in time-division duplexing systems, channel estimation can be performed by uplink pilot transmissions~\cite{M10}; CSI acquisition overhead scales with the number of users. As a result, the true value of $\eta^*$ may be smaller than the one obtained in this paper. However, our results show that \emph{at most} $60\%$ of the available spatial resources can be exploited to maximize sum-rate; the remaining should be reserved to provide diversity order for the scheduled users. 


In Fig.~\ref{fig:sum_rate5}, the optimal loading factor is obtained under ZF beamforming and equal power allocation assumptions. Although the obtained $\eta^*$ is consistent under various system parameters, we note that its value does depend on the chosen transmission strategy. To illustrate this point, we solve a sum-rate maximization problem using the weighted minimum mean square error (WMMSE) algorithm as proposed in~\cite{SRLH11} for a network consisting of seven hexagonal clusters. The WMMSE approach jointly optimizes user scheduling, beam directions and power assignments as to maximize sum-rate across the entire network. Note that such an algorithm would also require inter-cluster CSI; hence, it is more difficult to implement in practice. Without having to choose the loading factor a priori, the WMMSE algorithm implicitly schedules the right number of users during each time-slot to maximize the sum rate. The per-BS ergodic sum rate and the optimal loading factors under WMMSE are listed in Table~\ref{table:wmmse} for different values of $\bar{B}$. As seen from the table, the optimal loading factor is $\eta^* \in \LSB 0.5,0.55 \RSB$ in all considered scenarios, which is not far off from the optimal loading factor obtained from our analysis for ZF beamforming.

\begin{table}[!t]
\caption{Per-BS Ergodic Sum Rate and Optimal Loading factor Using WMMSE}
\centering
    \begin{tabular}{|c|c|c|}
    \hline 
    Average number of BSs & Per-BS ergodic sum rate (bits/sec/Hz) & $\eta^*$\\  \hline
    $2$ & $9.70$ & $0.52$\\ \hline
    $3$ & $10.36$ & $0.52$\\ \hline
    $4$ & $10.90$ & $0.53$\\ \hline  
         \end{tabular}
    \label{table:wmmse}
\end{table}

\subsection{Asymptotic Per-BS Ergodic Sum Rate with $\eta = 1$ and $\eta <1$}

So far, we have obtained the optimal loading factor that maximizes the per-BS ergodic sum rate. This section now shows that realizing the performance gain of network MIMO systems is a strong function of the loading factor $\eta$. To do this, we consider an asymptotic regime where the cluster size goes to infinity, i.e., $R_c \rightarrow \infty$, while $\eta$ is fixed. The main result of this section is the following upper bound on the asymptotic per-BS ergodic sum rate of the network MIMO system:

\begin{thm}
\label{asymp_per_cell}
Under Approximations~\ref{mom_match_app}-\ref{newapp}, the asymptotic per-BS ergodic sum rate of a network MIMO system operating with ZF beamforming and equal power assignment is upper bounded by
\begin{equation}
\lim_{R_c \rightarrow \infty}R_{cell} \leq \eta M \log_2 \LB 1 + \frac{2\rho M \lambda\pi d_o^2 \LB 1 - \eta \RB}{\Gamma \LB \alpha - 1 \RB \LB \alpha - 2 \RB} \RB. \nonumber
\end{equation}
\end{thm}

\begin{IEEEproof}
See Appendix~\ref{sec:Asymp}.
\end{IEEEproof}

\begin{rem}
The upper bound is obtained by analyzing the signal power alone, while ignoring inter-cluster interference. The main implication of Theorem~\ref{asymp_per_cell} is that, even when CSI acquisition overhead and backhaul constraints are ignored, the per-BS ergodic sum rate of a fully-loaded system approaches zero as the cluster size grows to infinity. In other words, increasing the cooperating cluster size in fact degrades system performance. However, when $\eta < 1$, the per-BS ergodic sum rate is upper bounded by a non-zero constant.
\end{rem}

To explain this asymptotic result, we note that the signal power for each user is a function of $\LB a \RB$ the channel strength, and $\LB b \RB$ the dimension of the vector subspace available to each user for beamforming, and in particular the ratio of the dimension of the beamforming subspace $\zeta$ to the dimension of the entire vector space. A channel with larger magnitude can provide larger signal power. Similarly, larger signal power can be harvested for a user if its beamforming subspace occupies a larger portion of the entire vector space. In the following, we observe that the channel strength is essentially a constant as cluster size grows to infinity, while the fraction of vector space available to each user for beamforming depends very much on the loading factor. In fact, this fraction goes to zero if $\eta = 1$. This explains the asymptotic behavior noted above.

First, consider the channel strength:

\begin{prop} \label{chan_ub}
Under Approximation~\ref{mom_match_app}, for user $i$ located at distance $d$ from the cluster center, the expected strength of the network MIMO channel is upper bounded by
\begin{equation}
\EE \LSB \| \g_{il} \|^2 \RSB \leq \frac{2 M\lambda \pi d_o^2}{\LB \alpha - 1 \RB \LB \alpha - 2 \RB}. \label{eq:chan_strength}
\end{equation}
\end{prop}
\begin{IEEEproof}
See Appendix~\ref{sec:chan_ub}
\end{IEEEproof}
Thus, although the dimension of the network MIMO channel increases as a larger number of BSs participate in forming a cluster, channel strength saturates. In other words, as $R_c \rightarrow \infty$, the spatial dimensions provided by the BSs located relatively far from a given user effectively do not contribute to the channel strength. 

The second key consideration is the ratio of the dimension of each user's beamforming subspace $\zeta = MB \LB 1 - \eta \RB + 1$ to the entire vector space $MB$. As the cluster size increases, we have
\begin{equation}
\lim_{R_c \rightarrow \infty} \frac{\zeta}{MB} = \left\{
\begin{array}{l l} 0 & \quad \text{if $\eta = 1$} \\ 1 - \eta & \quad \text{if $\eta < 1$}
\end{array} \right.. \nonumber
\end{equation}
Note that this ratio appears in the shape parameter of the Gamma distributions in our analysis of network MIMO systems.

Together, these two observations explain the asymptotic result of Theorem~\ref{asymp_per_cell}. In a fully-loaded system with $\eta = 1$, the channel strength does not grow as the cluster size increases, while $\frac{\zeta}{MB}$ approaches zero. The signal power therefore must go to zero as cluster size grows to infinity. In a partially-loaded system, although distant BSs do not enhance the channel strength, asymptotically, a non-zero fraction of the vector space is reserved for choosing the beamforming vector. Therefore, some of the signal power can be retained.

The asymptotic analysis presented in this section illustrates that in designing network MIMO systems with ZF beamforming and equal power assignment, it is crucial to distinguish between a fully-loaded and a partially-loaded system. Increasing the cluster size in a fully-loaded system degrades system performance.

Finally, we comment that the above conclusion depends critically on the assumption of ZF beamforming with fixed scheduling. In the remaining of this section, we numerically examine whether a better user scheduling and a better beamforming scheme can improve the asymptotic upper bound. To do this, we consider a network MIMO system with ZF beamforming and semi-orthogonal user scheduling, and a similar system with regularized ZF (RZF) beamforming and random user scheduling. From Theorem~\ref{asymp_per_cell}, the asymptotic upper-bound on the per-BS ergodic sum rate is achieved by ignoring the inter-cluster interference. Therefore, we consider only a single isolated cluster of BSs.
 
\begin{figure}[t!]
\centering
\includegraphics[scale = 1.03]{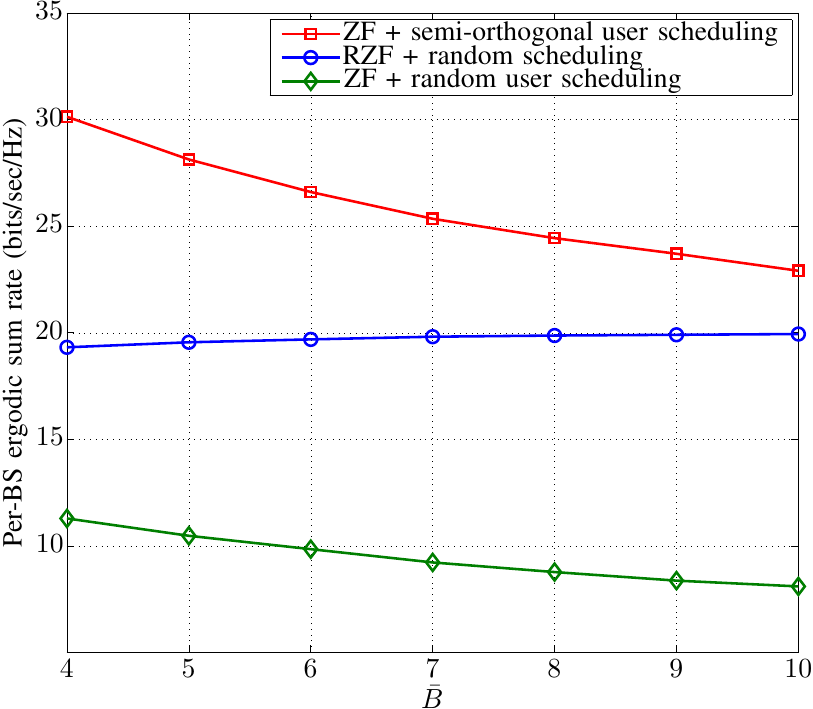}
\caption{Per-BS ergodic sum rate in an isolated cluster with $\eta = 1$ under different beamforming and user scheduling schemes.}
\label{fig:scheduling}
\end{figure}

The implementation of the semi-orthogonal user scheduling is based on the algorithm proposed in~\cite{YG06}. In each case, the number of potential users is assumed to be 100 times larger than the total number of transmit antennas in a cluster. As seen from Fig.~\ref{fig:scheduling} at $\eta = 1$, the per-BS ergodic sum rate of network MIMO approaches zero even with semi-orthogonal scheduling. The reason is that even with a large, but finite, number of potential users, the selected users cannot be made truly orthogonal. Therefore, the asymptotic result of Theorem~\ref{asymp_per_cell} still applies here.

Further, we consider RZF beamforming with random user scheduling. To design RZF beams, the regularization factor is set to $\frac{1}{\rho}$. As depicted in Fig.~\ref{fig:scheduling}, the per-BS ergodic sum rate no longer decreases as the cluster size increases at $\eta = 1$. Therefore, employing RZF can improve the asymptotic per-BS ergodic sum rate of a fully-loaded network MIMO system. 

\section{Network MIMO Performance as a Function of Cluster Size} \label{sec:cluster_size}
\subsection{Per-BS Ergodic Sum Rate at Different Cluster Sizes}
This section investigates the impact of cluster size on the per-BS ergodic sum rate of a network MIMO system evaluated at $\eta^* = 0.6$. Similar to the preceding section, the analytical results are obtained using Eqn.~\eqref{eq:sub_eqn}, and then compared with simulations under the system parameters as listed in Table~\ref{table:parameters}.

\begin{figure} 
\centering
\includegraphics[scale = 1.03]{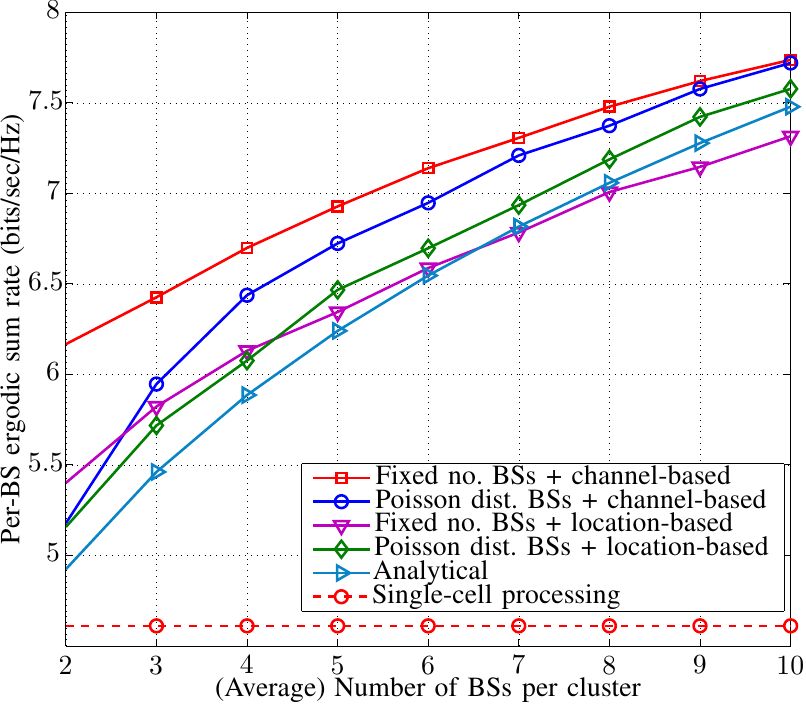}
\caption{Per-BS ergodic sum rate evaluated at $\eta^*$ for single-cell processing, and for network MIMO system with various (average) cluster sizes, with fixed number or Poisson distributed BSs in each cluster, and with channel-based and location-based user association schemes.}
\label{fig:rate_per_cell}
\end{figure}

This section considers cluster sizes of up to $10$ BSs, (beyond which numerical simulations would be too time-consuming.) The analysis of this paper assumes Poisson distribution of BS locations. To evaluate the effect of random BS deployments, we also consider a deployment scenario where each hexagonal cluster contains a fixed number of BSs that are uniformly distributed within the cluster region. Further, the analysis of this paper assumes a location-based user-to-cluster association scheme, i.e., only users located within the cluster region are associated with the cluster. In the numerical simulation, we also consider a more realistic user-to-cluster association scheme where each user is associated with a cluster that provides the largest average composite-channel gain. In particular, using the distribution of the network MIMO channel strength given in~\eqref{eq:intended_chan}, user $i$ is associated with cluster $l$ if
\begin{equation}
\sum_{b = 1}^{B_l} \beta_{ilbl} > \sum_{m = 1}^{B_j} \beta_{ijmj}~\forall{j \neq l}. \nonumber
\end{equation}
Finally, we compare network MIMO performance with a conventional single-cell processing scheme where users are associated with their closest BSs; each BS independently serves its own scheduled users. In this case, for $\eta = 0.6$ and $M = 5$, each BS serves $K = 3$ users in the downlink. The analytical and simulation results are plotted in Fig.~\ref{fig:rate_per_cell}.

Fig.~\ref{fig:rate_per_cell} shows that our analysis agrees with the simulation results very well and that considerable performance improvement in per-BS ergodic sum rate can be obtained as cluster size of a network MIMO system increases. Specifically, about a $70\%$ sum-rate improvement can be realized in a network MIMO system with $\bar{B} = 10$ as compared to a single-cell processing scheme of conventional systems. Fig.~\ref{fig:rate_per_cell} also shows that as the average cluster size increases, the performance of a network MIMO system with Poisson distributed BSs approaches that of a comparable system with a fixed number of BSs per cluster. Finally, as shown in Fig.~\ref{fig:rate_per_cell}, with both fixed and Poisson distributed numbers of BSs per cluster, as expected, a network MIMO system with channel-based user association outperforms the location-based association scheme. It should be noted that the analysis of network MIMO systems with channel-based user association would have been much more complicated than the location-based user association assumed in this paper. However, Fig.~\ref{fig:rate_per_cell} illustrates that the location-based analysis in this paper already provides a reasonable approximation to the per-BS ergodic sum rate for a network MIMO system employing a channel-based user-to-cluster association scheme.


\subsection{Asymptotic Per-BS Ergodic Sum Rate}
\begin{figure} 
\centering
\includegraphics[scale = 1.03]{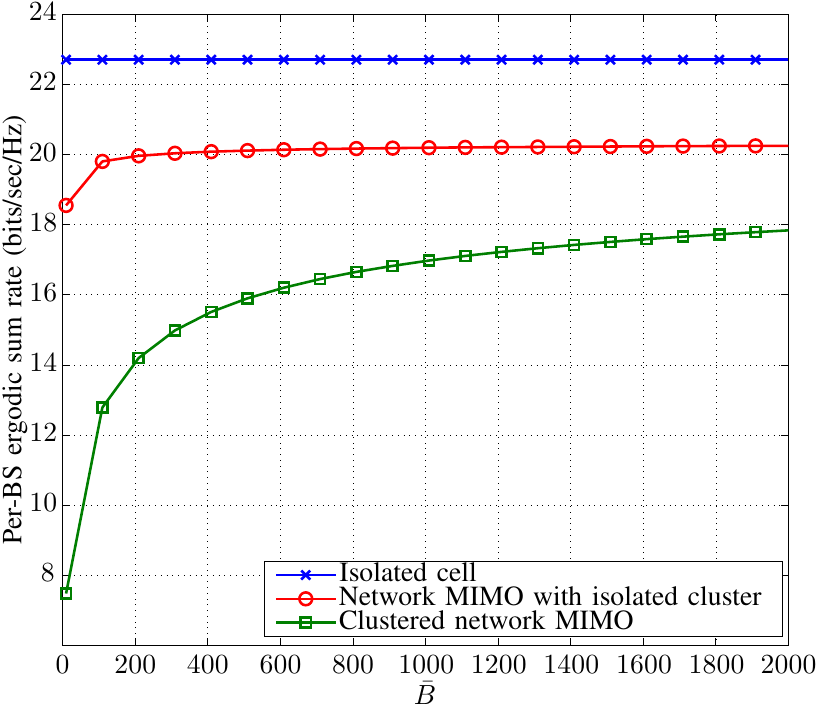}
\caption{Per-BS ergodic sum rate evaluated at $\eta^*$ for an isolated cell, a network MIMO with an isolated cluster, and a clustered network MIMO system under various average cluster sizes.}
\label{fig:IsolatedvsClustered}
\end{figure}

Our next objective is to understand the fundamental limitations of network MIMO systems by comparing the analytical expression of the network MIMO ergodic sum rate in our analysis (i.e.,~\eqref{eq:sub_eqn}) to two reference systems: $\LB a\RB$ a network MIMO system with an isolated cluster of $\bar{B}$ BSs and no out-of-cluster interference; $\LB b\RB$ a single isolated cell with one BS and no interference. The comparison with system $\LB a\RB$ above quantifies the effect of inter-cluster interference. The comparison with system $\LB b\RB$ above quantifies the signal strength penalty of cooperating across multiple BSs as compared to transmitting from a single BS alone. The numerical comparison of the three systems is shown in Fig~\ref{fig:IsolatedvsClustered}.

First, we observe that the ergodic sum rate of the network MIMO system in our analytical expression (which takes out-of-cluster interference into account) is always inferior to an isolated cluster of network MIMO systems with no out-of-cluster interference, \emph{even as the cluster size becomes very large}. This is because despite zero-forcing eliminating intra-cluster interference, inter-cluster interference always exists, when cluster size is finite. Such out-of-cluster interference severely affects users closer to the cluster edge. This illustrates that the performance of network MIMO systems with disjoint clustering is fundamentally limited by the inevitable out-of-cluster interference. We note that this conclusion agrees with the information theoretical study of cooperation gain (albeit under a different model) by Lozano \emph{et al.}~\cite{LHA13}.

Second, we observe that even without out-of-cluster interference, the performance of network MIMO systems does not approach that of an isolated cell with a single BS. This illustrates the fact that cooperating across multiple BSs introduces a significant signal power penalty. This penalty is due to the disparity in the distances between the user and the set of cooperating BSs in a network MIMO system. 

Mathematically, we can explain the origin of this penalty as follows. Considering two Gamma random variables $X_1$ and $X_2$, if both the shape and scale parameters of $X_1$ are larger than those of the $X_2$, then it is guaranteed that $X_1$, in first-order stochastic sense, is larger than $X_2$ (refer to~\cite{HYA14,DKA13} for more details.) The scale parameter of the Gamma distribution given by~\eqref{eq:signal_dist} improves if a user is served by only its closest BS rather than the group of cooperating BSs. Hence, the scale parameter of the signal power distribution is larger in an isolated cell system. In an isolated cell system, the shape parameter of the signal power distribution is $M - K + 1$. In a network MIMO system, the shape parameter is in the range of $M \LB 1 - \eta \RB + 1/B_l \leq k_{il} \leq MB_l \LB 1 - \eta \RB + 1$; the lower-bound is achieved when only one of the path-loss components is non-zero, and the upper-bound is achieved in a zero-probability event that the channels are isotropic, i.e., all path-loss components are equal. Although the shape parameter is not necessarily smaller in a network MIMO system as compared to the isolated cell system, the per-BS ergodic sum rates plotted in Fig.~\ref{fig:IsolatedvsClustered} illustrates that the received signal powers are in fact statistically weaker in a network MIMO system.



Taken together, due to both the signal power penalty incurred by joint data transmission from distributed BSs, and the presence of inter-cluster interference, the achievable per-BS ergodic sum rate of a network MIMO system does not reach that of an isolated cell. Even when CSI is available at no cost, Fig.~\ref{fig:IsolatedvsClustered} shows that a network MIMO system with $2000$ cooperating BSs only achieves about $80\%$ of the performance of an isolated cell without interference.

%

\begin{figure} 
\centering
%
%
%
%
%
%
%
%
%
%
%
\includegraphics[scale = 1.03]{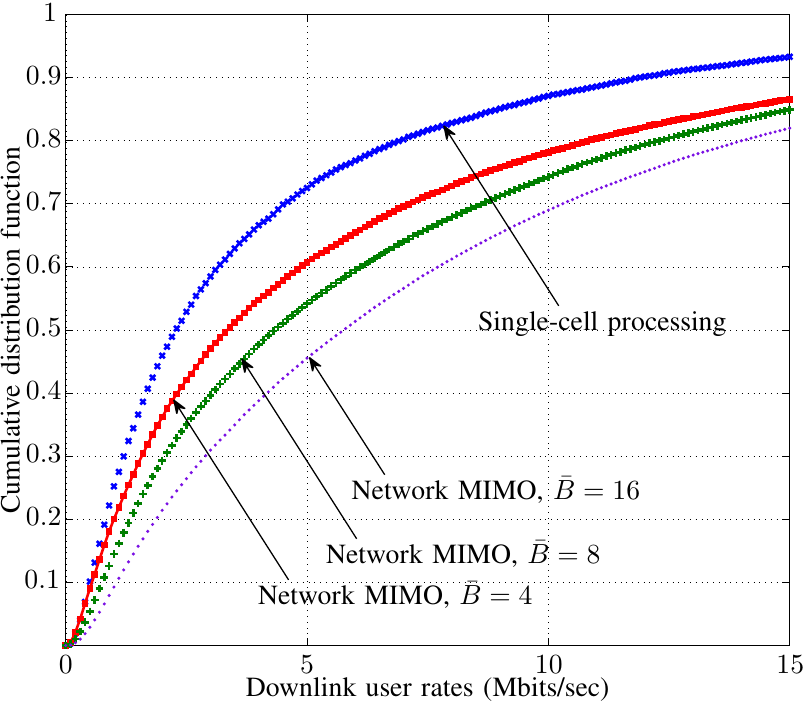}
\caption{Cumulative distribution function of the downlink user rates evaluated at $\eta^*$ for single-cell processing, and for network MIMO system under various average cluster sizes, Poisson distributed BSs in each cluster, and channel-based user association scheme.}
\label{fig:cdf_netMIMO}
 \end{figure}

Nevertheless, it should be emphasized that network MIMO does provide significant benefit as compared to single-cell processing treating out-of-cell interference as noise. Fig.~\ref{fig:cdf_netMIMO} shows the cumulative distribution function of downlink user rates of network MIMO systems with various cluster sizes. The density of users is set to be twenty times 
the density of BSs. A round-robin scheduler is used. Simulations show
that significant rate improvements can be realized in a network MIMO system as compared to the single-cell processing scheme of conventional systems. As an example, a network MIMO system with $\bar{B}=16$ doubles user rates as compared to single-cell processing scheme.


\section{Concluding Remarks}\label{sec:conc}
This paper develops an analytical tool to study the cooperation gain of network MIMO systems with disjoint clusters operating under ZF beamforming and equal power assignment. The proposed model takes the spatial statistics of user and BS locations, and random fading realizations into account. Specifically, we derive an accurate and computationally efficient per-BS ergodic sum rate expression using tools from stochastic geometry. This expression is used to investigate the performance of network MIMO systems as a function of loading factor and cooperating cluster size.

This paper shows that realizing the cooperation gain of network MIMO systems depends crucially on the loading factor. The optimal loading factor that maximizes the per-BS ergodic sum rate is $\eta^* \simeq 0.6$. Thus, at most $60\%$ of the spatial dimensions available in each cluster should be used to provide multiplexing gain; the remaining fraction of the spatial resources should be reserved to provide diversity for the scheduled users. 

The analysis of this paper also enables us to illustrate that the performance of network MIMO systems is fundamentally limited by the signal power penalty due to joint signal transmission from a disparity of BSs and by significant out-of-cluster interference. Regardless of the cluster size, network MIMO does not achieve the ergodic sum rate of an isolated cell. Nevertheless, network MIMO does provide significant improvement as compared to uncoordinated single-cell processing.

\appendices

\section{Proof of Theorem~\ref{ergodic_cap}} \label{sec:app_cap}
\begin{IEEEproof}
In cluster $l$, $K_l = \eta MB_l $ users are randomly chosen during each downlink transmission time interval. Therefore, the per-BS ergodic sum rate in nats per second per Hertz is given by
\begin{align}
R_{cell} &= \eta M \EE_{\Phi,\h} \LSB \int_{0}^{R_c} \log \LB 1 + \frac{\gamma_{il,d}}{\Gamma} \RB f_d \LB d \RB \mr{d}d \RSB \label{eq:per_cell1} \\
& = \eta M  \int_{0}^{R_c} \underbrace{\EE_{\Phi,\h} \LSB \log \LB 1 + \frac{\gamma_{il,d}}{\Gamma} \RB \RSB}_{R_{il,d}} f_d \LB d \RB \mr{d}d \label{eq:per_cell2} 
\end{align}
where $f_d \LB d \RB$ is given in~\eqref{eq:dist_dist}. The second equality is obtained by noting that the integrand in~\eqref{eq:per_cell1} is non-negative. Based on the Fubini theorem therefore the order of the expectation and integration can be exchanged.

The per-BS ergodic sum rate is a function of $R_{il,d}$; we therefore start the proof by characterizing $R_{il,d} $. To progress on this front, using the following equality~\cite{H08} 
\begin{equation}
\log \LB 1 + x \RB = \int_{0}^{\infty} \frac{e^{-t}}{t} \LB 1 - e^{-xt}\RB \mr{d}t \label{eq:log_exp}
\end{equation}
$R_{il,d}$ can be written as
\begin{align}
\EE_{\Phi,\h} & \LSB \int_{0}^{\infty} \frac{e^{-t}}{t} \LB 1 - \exp \LB \frac{-t \rho \lv \g_{il}\htp \w_{il}\rv^2}{\sum_{ j \ne l} \sum_{k} \rho \Gamma \lv \f_{ilj}\htp \w_{kj}\rv^2  + \Gamma} \RB \RB \mr{d}t \RSB \nonumber\\
\stackrel{\LB a \RB}{=}  &\int_{0}^{\infty} \frac{e^{-z\Gamma}}{z} \underbrace{\EE_{\Phi_I,\h} \LSB \exp \LB -z \rho \Gamma \sum_{ j \neq l} \sum_{k} \lv \f_{ilj}\htp \w_{kj}\rv^2 \RB \RSB}_{\Mc_I} \nonumber \\
&\times  \LSB 1 - \underbrace{\EE_{\Phi_S,\h} \LSB \exp \LB -z\rho \lv \g_{il}\htp \w_{il}\rv^2 \RB \RSB}_{\Mc_S} \RSB \mr{d}z. \label{eq:mgf_eqn}
\end{align}
In $\LB a \RB$, we use the change of variables as $t = z \Gamma\LB \sum_{j \neq l} \sum_k \rho \lv \f_{ilj} \htp \w_{kj} \rv^2 + 1 \RB$. Since the integrand is non-negative, the order of expectation and integration operations can be exchanged. Further, we use the fact that $\Phi_S$ and $\Phi_I$ are statistically independent. Note also that~\eqref{eq:mgf_eqn} involves the moment generating functionals (MGF) of the aggregate interference power $\Mc_I$ and the signal power $\Mc_S$.

 For the interference component, we have that
\begin{align}
&\Mc_{I} \stackrel{\LB a \RB}{=} \EE_{\Phi_I,\psi} \LSB \exp \LB -z \rho \Gamma \sum_{m \in \Phi_I} \beta_{ilm} \psi_{ilm} \RB \RSB \nonumber \\
&\stackrel{\LB b \RB}{=} \EE_{\Phi_I} \LSB \prod_{m \in \Phi_I} \EE_{\psi} \LB e^{-z \rho \Gamma \beta_{ilm} \psi_{ilm}}\RB \RSB \nonumber \\
&\stackrel{\LB c \RB}{=} \EE_{\Phi_I} \LSB \prod_{m \in \Phi_I} \LB 1 + z\rho \Gamma \LB 1 + \frac{r_{ilm}}{d_o} \RB^{-\alpha} \RB^{-\eta M} \RSB \nonumber \\
&\stackrel{\LB d \RB}{=} e^{ \lambda \int_{0}^{2\pi} \int_{l_{\theta}}^{\infty} \LB \LB 1 + z\rho \Gamma \LB 1 + \frac{r}{d_o}\RB^{-\alpha}\RB^{-\eta M} -1 \RB r\mr{d}r\mr{d}\theta}. \nonumber
\end{align}
In $\LB a \RB$, we replaced the aggregate interference power by $\sum_{m \in \Phi_I} \beta_{ilm} \psi_{ilm}$. According to Approximation~\ref{intf_app2}, the two terms are equivalent in distribution. Since $\Phi_I$ and $\psi$ are independent, we obtain $\LB b \RB$. Relation $\LB c \RB$ follows from the MGF of a Gamma distribution with the parameters as given in~\eqref{eq:psi}. Using the probability generating functional of a homogeneous PPP with density $\lambda$ as given by~\cite{H12}
\begin{equation}
\EE \LB \prod_{\x \in \Phi} v \LB \x \RB \RB = \exp \LB -\lambda \int_{\RR^2} \LB 1 - v\LB \x \RB \RB \mr{d}\x\RB \label{eq:mgf_ppp}
\end{equation}
and converting the coordinates from Cartesian to polar, we obtain the final expression. The region of integration can be determined by noting that the distance between the interfering BSs and user $i$ in cluster $l$ depends on the angle $\theta$ as shown in Fig.~\ref{fig:sys_model}. In particular, for a given $\theta$, the distance lies in the range $[ l_{\theta},\infty )$, where 
\begin{equation}
l_{\theta} = \sqrt{R_c^2 - d^2 \cos^2 \LB \theta \RB} + d \sin \LB \theta \RB \label{eq:distance}
\end{equation}
denotes the distance between user $i$ located at distance $d$ from the center of a typical cluster $l$ and the cluster boundary as depicted in Fig.~\ref{fig:sys_model}.

Likewise, for the signal component, it follows that
\begin{align}
&\Mc_{S} \stackrel{\LB a \RB}{=} \EE_{\Phi_S,\tilde{\kappa}} \LSB \exp \LB -z \rho \sum_{b \in \Phi_S} \beta_{ilbl} \tilde{\kappa} \RB \RSB \nonumber \\
&\stackrel{\LB b \RB}{=} \EE_{\Phi_S} \LSB \prod_{b \in \Phi_S} \EE_{\tilde{\kappa}} \LB e^{-z \rho \beta_{ilbl} \tilde{\kappa}}\RB \RSB \nonumber \\
&\stackrel{\LB c \RB}{=} \EE_{\Phi_S} \LSB \prod_{b \in \Phi_S} \LB 1 + z\rho \LB 1 + \frac{r_{ilbl}}{d_o} \RB^{-\alpha} \RB^{-\varpi} \RSB \nonumber \\
&\stackrel{\LB d \RB}{=} e^{ \lambda \int_{0}^{2\pi} \int_{0}^{l_{\theta}} \LB \LB 1 + z\rho \LB 1 + \frac{r}{d_o}\RB^{-\alpha}\RB^{-\varpi} -1 \RB r\mr{d}r\mr{d}\theta}. \nonumber
\end{align}

In $\LB a \RB$, we replaced the signal power by the random variable of the same distribution as proposed in Approximation~\ref{sig_mom_match2}. Relation $\LB b\RB$ follows from the independence of $\Phi_S$ and $\tilde{\kappa}$. In $\LB c\RB$, the MGF of a Gamma random variable with a distribution function given in~\eqref{eq:sig_app2} is used. Here
\begin{equation}
\varpi = \frac{M\bar{B}  \LB 1 - \eta \RB + 1}{\bar{B}}. \nonumber
\end{equation}
Using~\eqref{eq:mgf_ppp}, and by converting the coordinates from Cartesian to polar, we have the final expression where the region of integration depends on the user location. Specifically, since user $i$ is at distance $d$ from the center of cluster $l$, its distance to the cluster boundary is in the range $\LSB 0,l_{\theta}\RSB$ for a fixed angle $\theta$, as depicted in Fig.~\ref{fig:sys_model}. 

By replacing $\Mc_I$ and $\Mc_S$ in~\eqref{eq:mgf_eqn}, and applying the following integral equations
\begin{align}
&\int \LB 1 + ax^{-b}\RB^{-k}x \mr{d}x = \frac{x^2}{2} {}_2F_1 \LB k,-\frac{2}{b};1 - \frac{2}{b};-ax^{-b}\RB \nonumber \\
&\int \LB 1 + ax^{-b}\RB^{-k} \mr{d}x =  x {}_2F_1 \LB k,-\frac{1}{b};1 - \frac{1}{b};-ax^{-b}\RB\nonumber
\end{align}
we derive the ergodic rate of user $i$ located at distance $d$ from the center of cluster $l$. Finally, by inserting the user ergodic rate expression in~\eqref{eq:per_cell2}, we obtain the per-BS ergodic sum rate expression for a network MIMO system. The proof is therefore complete.

\end{IEEEproof}


\section{Proof of Theorem~\ref{asymp_per_cell}} \label{sec:Asymp}
\begin{IEEEproof}
From~\eqref{eq:per_cell2}, we have
\begin{equation}
R_{cell} = \eta M \int_{0}^{R_c} R_{il,d} f_{d}\LB d \RB \mr{d}d \label{eq:cell_user}
\end{equation}
where $R_{il,d}$ is given in~\eqref{eq:cap}. To analyze the asymptotic per-BS ergodic sum rate, we therefore upper bound $R_{il, d}$ as follows:
\begin{align}
R_{il,d} &\stackrel{\LB a \RB}{\leq} \EE_{\Phi_S,\h} \LSB \log_2 \LB 1 + \frac{\rho \lv \g_{il}\htp \w_{il}\rv^2}{\Gamma} \RB \RSB \nonumber \\
& \stackrel{\LB b \RB}{\leq} \log_2 \LB 1 + \frac{\rho}{\Gamma} \EE_{\Phi_S,\h} \LSB \lv \g_{il} \htp \w_{il} \rv^2\RSB\RB \label{eq:upper_bound}
\end{align}
where $\LB a \RB$ follows by ignoring the inter-cluster interference. The rate expression is a concave function of the signal power. Relation $\LB b \RB$ therefore holds using the Jensen's inequality. 

From~\eqref{eq:upper_bound}, to characterize the upper bound, we need the expected signal power. The expected signal power of user $i$ is given by
\begin{align}
&\EE_{\Phi_S,\h} \LSB \lv \g_{il} \htp \w_{il} \rv^2 \RSB \stackrel{\LB a\RB}{=} \EE_{B_l} \LSB \EE_{\Phi_S,\kappa_b \vert B_l}  \LSB \sum_{b = 1}^{B_l} \beta_{ilbl} \kappa_b  \RSB  \RSB \nonumber \\
&\stackrel{\LB b\RB}{=} \EE_{B_l} \LSB \sum_{b = 1}^{B_l} \EE_{\Phi_S \vert B_l} \LSB \beta_{ilbl} \RSB \EE_{\kappa_b \vert B_l} \LSB \kappa_b \RSB \RSB  \nonumber \\
&\stackrel{\LB c\RB}{=} \EE_{B_l} \LSB \frac{MB_l \LB 1 - \eta \RB + 1}{B_l} \EE_{\Phi_S \vert B_l} \LSB  \sum_{b = 1}^{B_l} \beta_{ilbl} \RSB \RSB \nonumber \\
&\stackrel{\LB d\RB}{=}\EE_{B_l}\LSB \LB MB_l \LB 1 - \eta \RB + 1 \RB \EE_{r} \LSB \LB 1+\frac{r}{d_o} \RB^{-\alpha} \RSB \RSB.\label{eq:exp_over_B}
\end{align}
Conditioned on the number of BSs in cluster $l$, the signal power is distributed as in Approximation~\ref{sig_mom_match1}. Using the law of total expectation, $\LB a\RB$ follows. In $\LB b\RB$, we use the fact that $\Phi_S$ and $\kappa_b$ are independent. According to Approximation~\ref{sig_mom_match1}, $\kappa_b$ is a Gamma random variable. Note that if $X \sim \Gamma \LB k,\delta \RB$, then $\EE \LB X \RB = k\delta$. Hence, $\LB c\RB$ holds. Conditioned on having $B_l$ BSs within the cluster, the distributions of their locations are independent and uniformly distributed~\cite{H12}; therefore, $\LB d\RB$ follows.
In $\LB d \RB$, $r$ denotes the distance between a uniformly distributed BS and user $i$ located at distance $d$ from the cluster center.

The expectation of the path-loss component can be computed as
\begin{align}
&\EE_r \LSB \LB 1 + \frac{r}{d_o} \RB^{-\alpha} \RSB \stackrel{\LB a \RB}{\leq} \frac{2}{R_c^2}  \int_{0}^{R_c} r \LB 1+ \frac{r}{d_o}\RB^{-\alpha}\mr{d}r  \nonumber\\
& \leq \frac{2d_o^2}{R_c^2 \LB \alpha - 1 \RB \LB \alpha - 2\RB}.  \label{eq:integral_path_loss}
\end{align}
The average path-loss component is larger when a user is located at the center, i.e., $d = 0$. In this case, the distribution of the distance between a user and a uniformly distributed BS is the same as~\eqref{eq:dist_dist}. The relation $\LB a\RB$ therefore follows.

By inserting~\eqref{eq:integral_path_loss} in~\eqref{eq:exp_over_B}, the expected signal power is given as
\begin{align}
&\EE_{\Phi_S,\h} \LSB \lv \g_{il} \htp \w_{il} \rv^2 \RSB \leq \EE_{B_l} \LSB \frac{2 d_o^2 \LB MB_l \LB 1 - \eta \RB + 1 \RB}{R_c^2 \LB \alpha - 1 \RB \LB \alpha - 2 \RB} \RSB \nonumber\\
& = \frac{2d_o^2}{R_c^2\LB \alpha - 1\RB \LB \alpha - 2\RB} + \frac{2M \lambda \pi d_o^2\LB 1 - \eta\RB}{\LB \alpha - 1 \RB \LB \alpha - 2\RB}. \label{eq:sig_ub} 
\end{align}

Therefore, from~\eqref{eq:upper_bound}, the user ergodic rate is upper bounded by
\begin{equation}
R_{UB} = \log_2 \LB 1 +  \frac{2\rho d_o^2}{\Gamma R_c^2\LB \alpha - 1\RB\LB \alpha - 2\RB} + \frac{2\rho  M \lambda \pi d_o^2\LB 1 - \eta \RB}{\Gamma\LB \alpha - 1 \RB \LB \alpha - 2 \RB} \RB. \nonumber
\end{equation}
Finally, if $\eta = 1$, we have that 
\begin{equation}
\lim_{R_c \rightarrow \infty}R_{UB} = 0. \nonumber 
\end{equation}
On the other hand, when $\eta < 1$, it follows that
\begin{equation}
\lim_{R_c \rightarrow \infty} R_{UB} = \log_2\LB 1 + \frac{2\rho M\lambda \pi d_o^2 \LB 1 - \eta \RB}{\Gamma \LB \alpha - 1 \RB \LB \alpha - 2 \RB}\RB. \nonumber
\end{equation}
By inserting these results in~\eqref{eq:cell_user}, the proof is complete.
\end{IEEEproof}

\section{Proof of Proposition~\ref{chan_ub}} \label{sec:chan_ub}
\begin{IEEEproof}
The expected network MIMO channel strength can be derived as
\begin{align}
&\EE_{\Phi_s,\h} \LSB \g_{il} \htp \g_{il} \RSB \stackrel{\LB a \RB}{=} M \EE_{B_l} \LSB \EE_{\Phi_s \vert B_l} \LSB \sum_{b = 1}^{B_l} \beta_{ilbl} \RSB \RSB \nonumber \\
&\stackrel{\LB b \RB}{=} M \EE_{B_l} \LSB B_l \EE_r \LSB \LB 1 + \frac{r}{d_o} \RB^{-\alpha} \RSB \RSB \nonumber\\
&\stackrel{\LB c \RB}{\leq} M \EE_{B_l} \LSB \frac{2d_o^2B_l}{R_c^2 \LB \alpha - 1\RB \LB \alpha - 2 \RB} \RSB \nonumber \\
&= \frac{2M\lambda \pi d_o^2}{\LB \alpha -1\RB \LB \alpha - 2 \RB}. \nonumber 
\end{align}

To obtain $\LB a\RB$, we use the distribution of network MIMO channel strength given by~\eqref{eq:intended_chan} and the law of total expectation. Conditioned on having $B_l$ BSs inside cluster $l$, they are uniformly and independently distributed. Therefore, $\LB b\RB$ follows. Using the upper-bound on the path-loss component in~\eqref{eq:integral_path_loss}, we have $\LB c\RB$. This obtains an upper-bound on the strength of network MIMO channel.
\end{IEEEproof}

\bibliographystyle{IEEEtran}
\bibliography{IEEEabrv,MyRef1}

\begin{IEEEbiography}[{\includegraphics[width=1in,height=1.25in,clip,keepaspectratio]{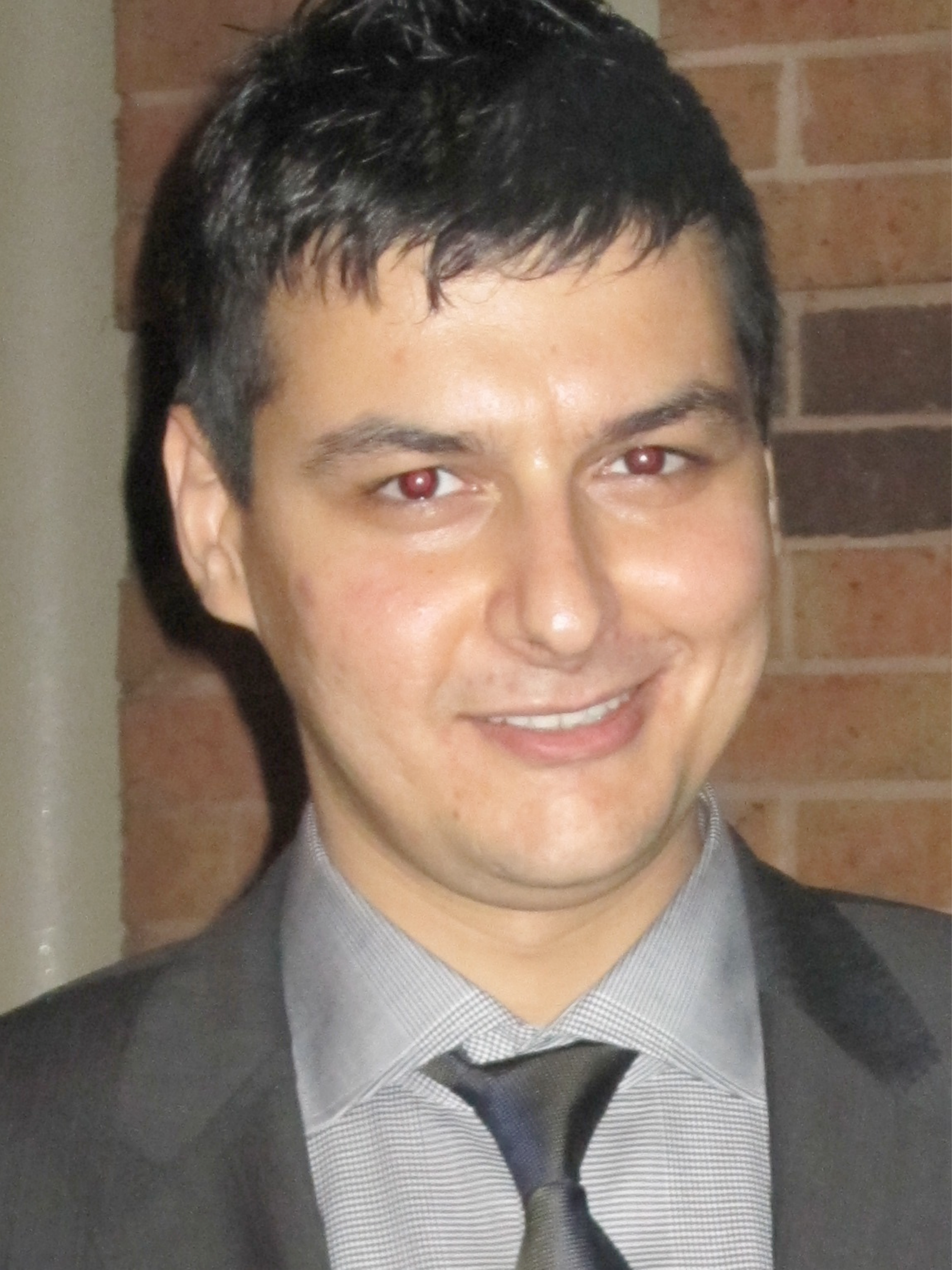}}]{Kianoush Hosseini}
(S'10-M'16) was born in Tehran, Iran. He received his B.Sc. degree in Electrical Engineering from Iran University of Science and Technology, Tehran, Iran in 2008 and M.A.Sc. and Ph.D. degrees in Electrical and Computer Engineering from University of Toronto, Toronto, ON, Canada in 2010 and 2015, respectively. Since 2015, he has been with Qualcomm Technologies Inc., San Diego, CA, USA. Dr. Hosseini is the recipient of the Edward S. Rogers Sr. Graduate Scholarship in 2008 and 2010, the Queen Elizabeth II Graduate Scholarship in Science and Technology in 2015, and Shahid U.H. Qureshi Memorial Scholarship in 2015. His main research interests include wireless communications, optimization theory, information theory, and distributed algorithms.
\end{IEEEbiography}

\begin{IEEEbiography}[{\includegraphics[width=1in,height=1.25in,clip,keepaspectratio]{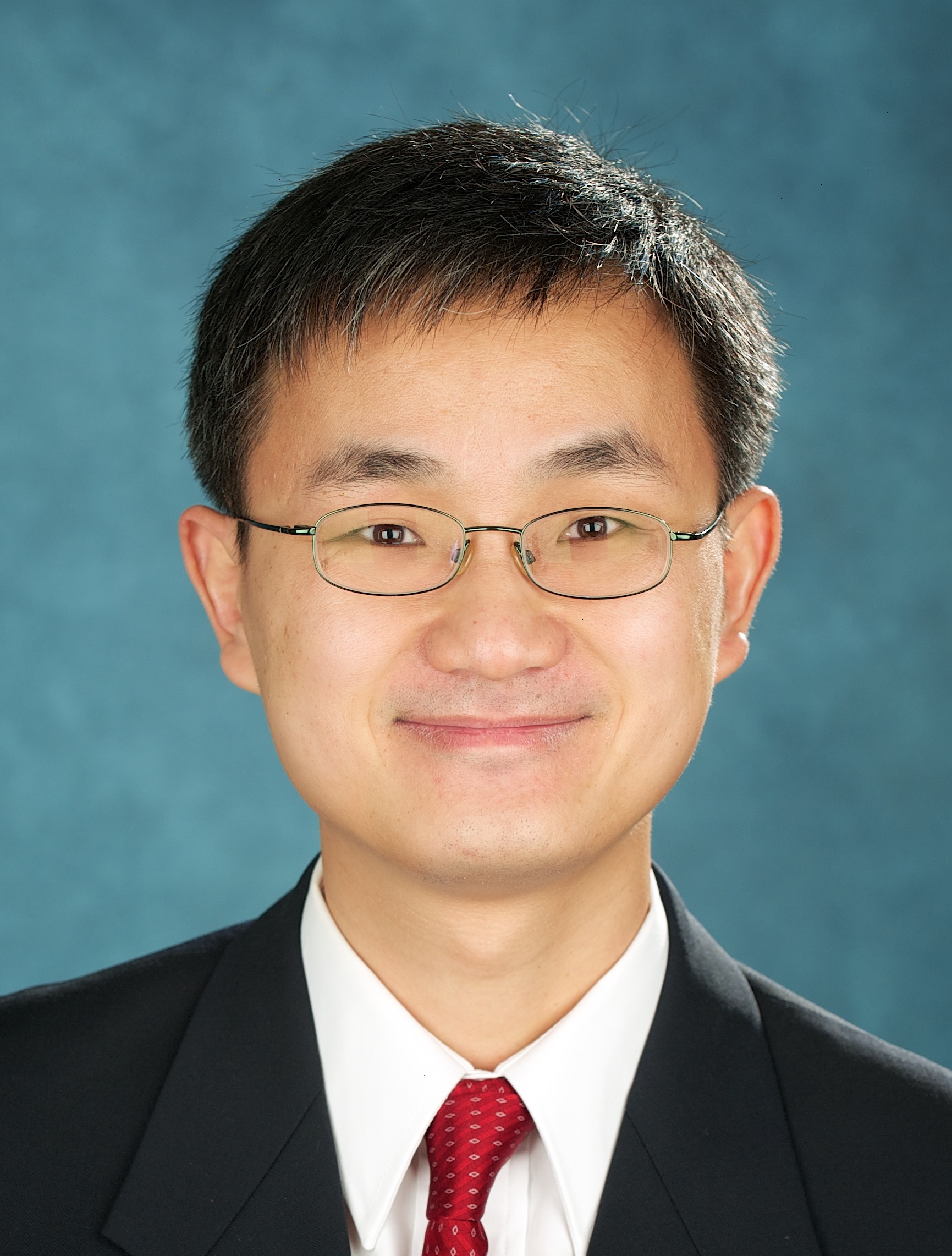}}]{Wei Yu}
(S'97-M'02-SM'08-FÕ14) received the B.A.Sc. degree in Computer Engineering and Mathematics from the University of Waterloo, Waterloo, Ontario, Canada in 1997 and M.S. and Ph.D. degrees in Electrical Engineering from Stanford University, Stanford, CA, in 1998 and 2002, respectively. Since 2002, he has been with the Electrical and Computer Engineering Department at the University of Toronto, Toronto, Ontario, Canada, where he is now Professor and holds a Canada Research Chair (Tier 1) in Information Theory and Wireless Communications. His main research interests include information theory, optimization, wireless communications and broadband access networks.

Prof. Wei Yu currently serves on the IEEE Information Theory Society Board of Governors (2015-17). He is an IEEE Communications Society Distinguished Lecturer (2015-16). He served as an Associate Editor for IEEE Transactions on Information Theory (2010-2013), as an Editor for IEEE Transactions on Communications (2009-2011), as an Editor for IEEE Transactions on Wireless Communications (2004-2007), and as a Guest Editor for a number of special issues for the IEEE Journal on Selected Areas in Communications and the EURASIP Journal on Applied Signal Processing. He was a Technical Program co-chair of the IEEE Communication Theory Workshop in 2014, and a Technical Program Committee co-chair of the Communication Theory Symposium at the IEEE International Conference on Communications (ICC) in 2012. He was a member of the Signal Processing for Communications and Networking Technical Committee of the IEEE Signal Processing Society (2008-2013). Prof. Wei Yu received a Steacie Memorial Fellowship in 2015, an IEEE Communications Society Best Tutorial Paper Award in 2015, an IEEE ICC Best Paper Award in 2013, an IEEE Signal Processing Society Best Paper Award in 2008, the McCharles Prize for Early Career Research Distinction in 2008, the Early Career Teaching Award from the Faculty of Applied Science and Engineering, University of Toronto in 2007, and an Early Researcher Award from Ontario in 2006. He is recognized as a Highly Cited Researcher by Thomson Reuters. 
\end{IEEEbiography}

\begin{IEEEbiography}[{\includegraphics[width=1in,height=1.25in,clip,keepaspectratio]{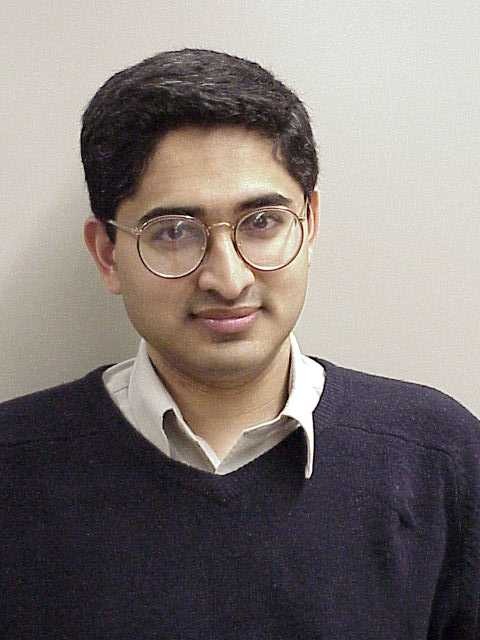}}]{Raviraj S. Adve}
(S'88-M'97-SM'06) was born in Bombay, India. He received
his B. Tech. in Electrical Engineering from IIT, Bombay, in 1990 and his
Ph.D. from Syracuse University in 1996. Between 1997 and August 2000, he
worked for Research Associates for Defense Conversion Inc. on contract with
the Air Force Research Laboratory at Rome, NY. He joined the faculty at the
University of Toronto in August 2000 where he is currently a Professor. Dr.
Adve's research interests include analysis and design techniques for
heterogeneous networks, energy harvesting networks and in signal processing
techniques for radar and sonar systems. He received the 2009 Fred Nathanson
Young Radar Engineer of the Year award.

\end{IEEEbiography}

\end{document}